\documentstyle[preprint,aps,epsfig]{revtex}

\newcommand {\Pomeron} {I\!\!P}
\newcommand {\dperp} {d_{\perp}}
\def\lsim{\mathrel{\rlap{\lower4pt\hbox{\hskip1pt$\sim$}}
    \raise1pt\hbox{$<$}}}                

\addtocontents{toc}{\protect\hspace{10mm}}

\begin{document}
\draft
\title{Electron-Nucleus Collisions at THERA}
\author{L. Frankfurt$^{1}$, V. Guzey$^{2}$, M. McDermott$^3$, 
and 
M. Strikman$^4$}
\address{$^1$ School of Physics and Astronomy, Tel Aviv University, 69978 
Tel Aviv, Israel}
\address{$^2$Special Research Centre for the Subatomic Structure of Matter (CSSM), Adelaide University, 5005, Australia.}
\address{$^3$Division of Theoretical Physics, Dept. of Math. Sciences, University of Liverpool, Liverpool L69 3BX, England.}
\address{$^4$Department of Physics, Pennsylvania State University,University Park, PA 16802, USA}
\date{\today}
\preprint{
\vbox{
\hbox{ADP-01-15/T450}
\hbox{LTH ??}
}}

\maketitle

\begin{abstract}

The nuclear option at THERA provides an ideal and unique opportunity to
investigate the black body limit (BBL) of high energy Deep Inelastic
Scattering of highly virtual photons off heavy nuclear targets and thereby
probe QCD in a new regime. At high enough energies, whichever hadronic
configuration the photon fluctuates into, the interaction at a given impact 
parameter with the heavy nuclear target will eventually reach its
geometrical limit corresponding to the scattering from a black disk. An
attractive feature of the BBL regime for a large nucleus is that the
interaction is strong although the coupling constant is small. Predictions
for  longitudinal and transverse distributions of the leading hadrons in
inclusive and diffractive channels and exclusive vector meson production
are found to be strikingly different in BBL and the leading twist
approximation. In particular, the multiplicity of leading hadrons in the
current fragmentation region is strongly suppressed, while the cross
section of diffractive vector meson production by longitudinally polarized
photons is $\propto 1/Q^2$. The  x and $Q^2$ ranges where BBL maybe
reached are calculated  for the interaction of color triplet and color
octet dipoles at central impact parameters. We conclude that for  heavy
nuclei THERA would reach deep into BBL for a wide range of $Q^2$. The
connection between hard diffraction and leading twist nuclear shadowing is
analysed. It is demonstrated that the current leading twist analyses of
the HERA diffractive data lead to similar and very large leading twist
shadowing for gluons. Several processes sensitive to the amount of nuclear
shadowing for gluons are discussed.

\end{abstract}

\section{Executive Summary}
\label{sec:nuc:summary}

The nuclear option at THERA provides an ideal and unique
opportunity to investigate the black body limit (BBL) of high
energy Deep Inelastic Scattering (DIS) of highly
virtual photons off heavy nuclear targets and thereby
probe Quantum Chromodynamics (QCD) in a new regime.
At high enough energies, whichever hadronic configuration
the photon fluctuates into, the interaction at given impact parameter
with the heavy nuclear target will eventually reach its geometrical
limit corresponding to the scattering from a black disk.
An attractive likely feature of the approach to the BBL regime for a large 
nucleus is that the interaction is strong although the coupling constant 
is small. Predictions for many observables are qualitatively different
within the BBL from those
of the standard leading twist DGLAP \cite{dglap} regime. The
most striking predictions
for inclusive observables include: (i) the nuclear structure
functions behave as
$F^{A}_{2} \propto \pi R_A^2 Q^2 \ln (1/4m_NR_Ax)$,
with a $1/x$ dependence slower than that predicted by DGLAP 
(this is in contrast with the nucleon case in which 
$F^{N}_{2} \propto Q^2 \pi R_N^2 (1+c_N \ln^{2} 1/x)\ln 1/x$
whose $x$ dependence is comparable with the DGLAP prediction, 
here $R_A,R_N$ are radii of nucleus and nucleon),
(ii) the ratio of nuclear and nucleon structure 
functions decreases with $1/x$, $F^{A}_{2} / F^{N}_2 \propto 1 / (1+c_N\ln^{2} 1/x)$.
Major new features for the final states in DIS
are: (i) a much softer leading hadron spectra with
enhanced production of high $p_t$ jets
in inclusive and diffractive processes, (ii) the inclusive
diffractive cross section reaches nearly half of the total cross section, 
(iii) a weakly $Q^2$-dependent parameter-free cross section for exclusive vector meson electroproduction.

It is possible that the black limit will be reached for certain
gluon-induced
DIS processes off nucleons
within THERA kinematics. However, nuclei have two very clear and
useful advantages over
nucleon targets. Firstly, because the target is
extremely
Lorentz contracted, the
partons of individual nucleons are piled up on one another. This
means that for a
given photon-target
centre of mass
energy the black limit
will be reached {\it much earlier}
 in nuclei. Secondly, scattering from the edge of nucleons, which
remains grey even
at very high energies, (and competes with the BBL contribution)
is heavily suppressed
for large nuclei. The purpose of the rest of this report is
to expand on the details
for the interested reader.

\section{Introduction}

The standard theoretical approach to inclusive Deep Inelastic
Scattering (DIS) of a
virtual-photon off a target, $\gamma^{\ast}(q) + T(p) \rightarrow X$, uses a
factorization theorem \cite{fact} to prove that in the
limit of large photon virtuality
($q^2 = - Q^2 \ll 0$), leading twist (LT) terms dominate in
the Operator Product Expansion
\cite{opeqcd} of the relevant hadronic matrix elements. This
corresponds to a factorization
of short and long distance contributions and leads to an attractive
physical picture of the virtual photon scattering off electrically charged
``point-like'' partonic
constituents within the target, with its associated
approximate
Bjorken \cite{bj} scaling (which
states that structure functions, for example,
$F_L$ and $F_2$, only depend on the
ratio $x = Q^2 / 2p\cdot q$). The application of the renormalization group to
perturbative Quantum Chromodynamics (pQCD), embodied in the
DGLAP \cite{dglap} evolution
equations, corrects this naive picture and leads to
logarithmic scaling violations.
These equations describe how the long-distance boundary conditions, the parton
distributions of the quarks and gluons, (and ultimately
$F_L$ and $F_2$) change logarithmically with $Q^2$.

It is now understood that this decomposition in terms of twists
should become inapplicable
in pQCD at sufficiently small $x$ (i.e. the higher twist (HT) terms,
which are formally
suppressed by positive integer powers of $1/Q^2$, become
numerically important). However,
in practice, it is very difficult to distinguish HT effects
in the inclusive structure
functions from uncertainties in the input parton distributions
for a nucleon target,
because DIS off the nucleon edge masks the importance
of the HT effects.
We will explain that the theoretical description of DIS off heavy nuclei
is significantly
simpler than that for a nucleon target. The energy and $Q^2$ dependence
of structure
functions of heavy nuclei, and specific properties of final states,
can serve as clear
signatures of a new regime, the black body limit (BBL) regime,
where significant
contributions to the dynamics of the strong interactions from
all twists are expected.

For the past decade, pQCD has been used extremely
successfully to describe inclusive DIS
processes (see e.g. \cite{pdf}). Many new phenomena
specific to small-$x$ physics have
been discovered experimentally and explained by pQCD. New QCD
factorization theorems for certain hard processes (such as diffractive scattering
\cite{coll}, exclusive vector meson production \cite{brod,cfs} and Deeply Virtual
Compton Scattering \cite{cf}) have been proven. These factorization theorems were used to
predict and describe
qualitatively new small-$x$ hard scattering phenomena. In
particular, hard exclusive
processes for both nucleon and nuclear targets, with
properties strikingly different
from those familiar from soft hadronic processes,
were predicted and observed (several
experimental \cite{AC} and theoretical \cite{FGSreview,MWUST,HEB}
reviews are available
in the literature). These include: (i) the complete
transparency of nuclear matter in
coherent pion diffraction into two jets off a nuclear
target (predicted in \cite{frs} and later observed in \cite{Ashery}) 
and in exclusive photoproduction of $J/\psi$ mesons off nuclei \cite{Sokoloff},
(ii) a fast increase with increasing energy of cross sections of
hard exclusive processes \cite{fs89prl}:
high $Q^2$ exclusive $\rho$, $\phi$, $J/\psi$ meson production
\cite{brod,fks1,fks2,ryskin,durrho}
and exclusive photoproduction of $J/\psi$ and $\Upsilon$ mesons
\cite{fms}, \cite{teub} and low-$x$ exclusive photon production in DIS
(Deeply Virtual Compton Scattering \cite{ffs}),
(iii) an almost universal and small slope of the $t$-dependence
of hard exclusive processes
(predicted in \cite{afs}),
 (iv) a small and process/scale
dependent rate of change of the slope of $t$-dependence
of exclusive processes with energy, $\alpha'$,
(v) confirmation that QCD is flavour
blind in hard (high $Q^2$) exclusive vector meson ($\rho$ and $\phi$ mesons)
electroproduction. These
theoretical and experimental
discoveries have moved the focus of investigations in
QCD to new frontiers.

A seminal and extremely important experimental observation is
the fast increase with $1/x$ of the proton structure
function, $F_2$,
at small $x$ discovered
by the H1 \cite{H1inc} and ZEUS \cite{ZEUSinc} collaborations at HERA.
The implied fast increase of the proton parton densities
raises several challenging questions for DIS at even higher energies:

\begin{itemize}
\item Will it remain sufficient to use the leading twist DGLAP
evolution equation (to a given accuracy in logarithms in $Q^2$) or
will it become necessary to resum \cite{bfkl,ciafaloni,forshawbook}
the large logs in energy that appear at each order in the perturbative
expansion (i.e. for the regime in which
$\alpha_s(Q^2) ~\ln(x_0/x) \sim {\cal O}(1)$) ?
\item Does diffusion in parton transverse momenta lead to a
breakdown of (collinear)
factorization for leading twist pQCD, and a non-trivial
mixing of perturbative and
non-perturbative
effects, in a kinematic regime in $Q^2$ which is assumed to be
safe at higher $x$ ?
\item Will the increase of the parton distributions lead to an
experimentally-accessible new pQCD regime, where the
coupling constant is small but the interaction is strong ?
If so, how can one distinguish between this new regime and the
standard one in which the LT DGLAP equations are applicable ?
\item Will the structure functions and cross sections of hard
exclusive processes continue to increase with energy or will their growth be
tamed to avoid violation of unitarity of the $S$-matrix
(applied to the hard
interactions
of the hadronic fluctuations of the photon) ?
\item Since the same QCD factorization theorems which lead to
successful description
of hard processes predict a rapid increase of HT effects with
increasing energy,
how important will the latter effects be for the
interpretation of the small-$x$ data ?
\end{itemize}

In the following we explain that one has a good chance to
answer these questions
by studying electron-nucleus collisions at THERA. We will
formulate several theoretical
predictions, based mostly on general properties of space-time
evolution of high-energy
scattering, which can serve as signals of new hard QCD phenomena.

DIS at high energies is most easily formulated in the target rest frame
(this approach has become known as {\it the dipole picture} for reasons
that will become apparent below). In this frame, the photon fluctuates
into a hadronic system (a quark anti-quark colour triplet {\it dipole}, to
lowest order in $\alpha_s$) a long distance,
$1/2 m_{T} x \gg R_{T}$, upstream of the target. 
This system interacts briefly with the target and eventually
the hadronic final state is formed. The formation time of the
initial state hadronic
fluctuations, and of the final state, is typically much longer
than the interaction time with the target.
This allows the process to be factorized into a wavefunction
for the formation of the initial state, an interaction cross
section with the target, and a wavefunction describing the formation of the
final state. In this frame, the standard LT result of pQCD, at
leading log accuracy, may be re-expressed using the following well-known
interaction cross-section
\cite{frs,fks1,fks2,mfgs}
for the scattering
of the $q {\bar q}$ dipole with the target (for an explicit
derivation see \cite{rad}):
\begin{equation}
{\hat \sigma}_{\mbox{{\scriptsize pQCD}}} (\dperp^2,x) = \frac{\pi^2}{3} \, \dperp^2 \, \alpha_s
({\bar Q^2}) \, x G^T (x,{\bar Q^2}) \, .
\label{eq:nuc:intr:eqshat}
\end{equation}
\noindent which is a generalization of the two-gluon exchange model of \cite{low} (see also references in \cite{MWUST}).
Here $\dperp$ is the transverse diameter of the dipole,
$xG^T$ is the
leading-log gluon distribution of the target sampled at a scale
${\bar Q^2} = \lambda/\dperp^2$, with $\lambda$ a logarithmic
function of $Q^2$.
There are of course corrections, involving more complicated partonic
configurations such as $q {\bar q} g$, which in general can not be described
(except in special circumstances) using the interactions of dipoles of a given
colour configuration. However, the description of small-$x$ processes in
terms of
an {\it interaction cross section} for the hadronic fluctuations of the photon
remains a useful one.

For heavy nuclear targets,  it is legitimate to neglect a logarithmic increase
of the effective  radius of the target  with increasing  energy. In this case there is, of course, a natural geometrical upper limit to the size of this
inelastic interaction cross section given approximately by the
transverse cross-sectional
area of the target:
\begin{eqnarray}
{\hat{\sigma}}_{\mbox{{\scriptsize pQCD}}} &\le \sigma_{\mbox{{\footnotesize black}}} & =
\pi R_T^2 \, .
\label{eq:nuc:intr:sigblack}
\end{eqnarray}

It seems plausible that small-$x$ physics of the next generation
of accelerators will be a
combination of two generic phenomena, {\it generalised colour transparency}
and {\it complete opacity} (absorption), which correspond to two different
kinematical limits. In one limit, when $x$ is fixed and
$Q^2\rightarrow \infty$,
hard processes exhibit numerous colour transparency phenomena, which can be
described in terms of various generalisations of the QCD
factorization theorem.
In another limit, when $Q^2$ is fixed and $x\rightarrow 0$, the
interaction cross
section increases with decreasing $x$ (increasing energy), which
leads to increased
absorption by the target. Hence, in this kinematical limit,
the approximation of
complete absorption, which could naturally be termed
``the black body limit (BBL)'',
seems to be a promising guide to predict new distinctive hard QCD phenomena.
Colour Transparency is a generic name used to describe the fact
that small colourless
configurations do not interact strongly with hadrons, which
is reflected by the
inclusion of the transverse diameter squared,
$\dperp^2$, in (\ref{eq:nuc:intr:eqshat}).
However, for a fixed dipole size (and neglecting for a moment an increase with
increasing energy of the radius of a target
which can be justified for heavy nuclear targets only), if the gluon density
of the target
continues to rise with increasing $1/x$, the interaction cross
section will eventually
reach its geometrical limit (cf. (\ref{eq:nuc:intr:sigblack})). This
is basically
what is meant by complete opacity.

The validity of the BBL can be justified for DIS processes off
heavy nuclear targets since contributions of colour transparency,
and various peripheral phenomena (which complicate studies of BBL
for nucleon targets) are suppressed.
In the BBL, the nuclear structure function $F_{2}^{A} (x,Q^2)$ is
predicted to behave as $R_A^{2} Q^2 \ln 1/x$ with decreasing $x$, which is
slower than the $x$-behaviour predicted by the DGLAP evolution equations.
Since the nucleon is ``gray'' rather than ``black'' as seen by the incoming
high-energy fluctuation, peripheral effects are important and contribute to
the total electron-nucleon DIS cross section
(these effects lead to an increase of the essential impact factors 
with increasing energy). 
The nucleon structure function (gluon distribution) behave as
$F_2^{N}\, (xG^{N}) \propto Q^2 R_N^2 \ln^{3} 1/x$
in the black body limit. In both the nuclear and nucleon cases,
the expression for the total cross sections is a direct generalization of the
Froissart unitarity limit \cite{Froissart} to DIS (see subsection~\ref{sec:nuc:black_nuc}).
One has to distinguish two features of these results. Firstly, the predicted
$Q^2$-independence of total cross sections grossly violates
Bjorken scaling and
serves as a very clean signal of the onset of the BBL regime.
Secondly, the predicted energy (or $x$) dependence of
the total electron-nucleon (but not electron-nucleus) cross section
(nucleon structure functions )
can probably be accommodated within the DGLAP LT pQCD framework
by a suitable subtle
re-tuning of the initial conditions (this has been the pattern
in the era of HERA).
Hence, it would be difficult to distinguish between
the DGLAP and black body approximations
and predictions thereof for the nucleon structure functions. A much more
promising way to identify the BBL regime is to investigate the difference 
between the $x$ dependence of nucleon structure function and the 
structure functions of heavy nuclei, as well as final states in DIS on 
nucleons and nuclei for which the BBL predicts several rather striking new phenomena.

By fixing $Q^2$ and decreasing $x$, one proceeds from
the DGLAP regime to the BBL regime.
Theoretical models attempting to quantify deviations from
the DGLAP approximation and
define the kinematical boundaries of the applicability of
the DGLAP equations agree that
the effects, which are responsible for the inapplicability
of the DGLAP dynamics, are
enhanced for nuclear targets by approximately a factor
of $A^{1/3}$ if the approximation is made that the HT effects 
and nuclear shadowing of parton distributions are a correction \cite{Mueller-Qiu}.
Neglecting non-perturbative nuclear shadowing of the gluon distribution, one 
can estimate the ratio of the critical $x$ at which the BBL regime sets in
for DIS on nuclei and nucleon, as
$x_{A} (\mbox{{\footnotesize BBL}}) / x_{N}(\mbox{{\footnotesize BBL}})
\approx (A/2)^{1/(3 n)}$ (the factor $n$ comes from the energy
dependence of the
nucleon structure function $F^N_{2}$
in the form $F^N_2\propto x^{-n}$).
The factor $1/2$ in this relation accounts for the difference between
the nucleon radius, $r_N=0.85$ fm, and a mean inter-nucleon distance 
in nuclei, $r_{NN}=1.7$ fm.

Note that the complication of taking non-perturbative nuclear shadowing in nuclear 
parton distributions into account slightly reduces the advantage of using nuclear 
targets over nucleon ones for studying higher twist effects.

Since the DGLAP QCD evolution equations are leading twist equations, any violations of
the DGLAP equation would signal a non-negligible role of higher twist effects.
Therefore, a unique feature of electron-nucleus collisions is that by studying
DIS on nuclear targets one can not only amplify the higher
twist effects but also
study them as a function of the target thickness (by varying
nucleon number $A$).

In order to summarize the above discussion, we would like
to re-emphasize that using
nuclear beams in THERA has significant advantages over using a
proton beam in studying new
QCD phenomena such as the perturbative regime of very high
parton densities and the
transition from the DGLAP regime to black body
scattering\footnote{Formally, the notion
of parton densities is defined in the leading twist approximation
only. Thus, when
discussing physics beyond the DGLAP approximation, one should
be cautious and explicit
in using the term ``parton density''.}. Since these phenomena
depend on the nucleon
number, $A$, measurements of various observables as a
function of $A$ will be necessary.
Fortunately this will be practical at THERA. The BBL interpretation
of many phenomena, including the difference between the $x$-dependence
of heavy nucleus and nucleon structure functions and some properties of final 
states, does not require knowledge of nuclear
effects. It is clear that a consistent and complete interpretation
of some experimental
data would require reliable information about traditional
nuclear effects at small $x$,
primarily nuclear shadowing. However, as we shall explain in detail in
Sect.~\ref{sec:nuc:shad}, using the profound connection
between nuclear shadowing
in inclusive electron-nucleus DIS and hard electron-proton
diffraction, the QCD
factorization theorem and modern fits to the diffractive data, one
can significantly
reduce uncertainties in the predictions of the
leading twist part of nuclear parton
densities.

The rest of this section is organized as follows. The black body
limit is discussed
in Sect.~\ref{sec:nuc:black_nuc}. We demonstrate that
predictions made in the
BBL are strikingly different from those obtained within the
DGLAP approximation,
especially for final states in DIS diffraction on nuclei. Precursors
of the BBL
can be studied by analysing the departure from the leading twist
DGLAP evolution
equation using the unitarity of the $S$-matrix
for hard interactions as a guide.
In Sect.~\ref{sec:nuc:unit}, we discuss the relevant
effects for a range of
nuclei at central impact parameters. It is found that our
predictions are sensitive
to the amount of leading twist nuclear shadowing, which is considered in
Sect.~\ref{sec:nuc:shad}. It is also demonstrated that in a wide range of
$x$ and $Q^2$, leading twist shadowing dominates over higher twist shadowing,
used frequently in eikonal-type models.
After briefly considering experimental requirements in
Sect.~\ref{sub:exp},
we conclude in Sect.~\ref{sec:nuc:conc}.

\section{The black body limit and its signals in DIS final states}
\label{sec:nuc:black_nuc}

The leading twist (LT) approximation of perturbative QCD
successfully describes DIS of photons on hadronic targets \cite{pdf}
and predicts a rapid increase
\footnote{This can be either introduced by hand at the initial evolution scale or  
generated by QCD evolution \cite{r74}.}
of structure functions with
increasing energy (decreasing $x$).
Within the LT approximation, there is no mechanism which would
slow down or tame the rapid growth of the structure functions
\footnote{One can show by direct pQCD calculations that the taming
(shadowing) of parton densities at small $x$ considered in
eikonal and parton recombination models is a higher twist effect.}.

However, it is clear that a rapid power-like growth of the structure
functions (at a given impact parameter) cannot continue forever
\footnote{At the same time, conventional nucleon structure functions
integrated over
all impact parameters may increase with increasing energy even faster
(according to the $\propto \ln^{3}1/x$ law) than predicted by DGLAP.}
and, hence, the LT approximation must be violated at very small $x$.
Therefore, some higher-twist mechanism is required to explain the taming
of the structure functions. A practical and almost model-independent
approach to the
taming of the structure function is based on the analysis of
the unitarity of the $S$-matrix
for the interaction of spatially small quark-gluon wave packets.

\subsection{Unitarity and the black body limit}
\label{sec:nuc:black_nuc:bbl}

The most direct way to understand the constraints which are imposed
by unitarity of the scattering matrix (for hadronic fluctuations of
the virtual photon) on the
nucleon and nucleus structure functions
is to use the {\it impact-parameter representation} for the scattering
amplitude.
As explained in the introduction, in the target rest frame the
incoming photon interacts with the target via its partonic fluctuations.
Since the interaction time is much shorter than the life-time of
the fluctuations at small $x$, the DIS amplitude can be factorised
in three factors: one describing the formation of the fluctuation, another --
the hard interaction with the target, and a final factor describing the
formation of the hadronic final state. The $q {\bar q}$ dipole is the dominant
fluctuation at short transverse distances, and we consider its
interaction primarily
in what follows.

Within this high-energy factorization framework the structure functions,
$F_L$ and $F_T$, may be written in the following simple way
\begin{equation}
F_{L,T} (x,Q^2)=\int dz\,d^2 \dperp |\psi^{\gamma}_{L,T} (z,\dperp^2,Q^2)|^2
{Im\,A(s,t=0,\dperp^2)\over s}  \, ,
\label{eq:nuc:black_nuc:bbl:1a}
\end{equation}
where $\psi^{\gamma}_{L,T} (z,\dperp^2,Q^2)$ are the light-cone wavefunctions
of the longitudinally and transversely polarised virtual photon,
respectively,
$z$ is the photon momentum fraction carried by one of
the dipole constituents, $s$ is the invariant
energy of the dipole-target system ($s = (P + q)^2$ for
the $q {\bar q}$ dipole),
$t \approx -|\vec{q^{'}}_{\perp}|^2 $
is the momentum transfer and $\dperp$ is the dipole's transverse diameter.

The dipole-target scattering amplitude $A(s,t,\dperp^2)$ can be expressed via
the corresponding amplitude, $f(s,b,\dperp^2)$, in
impact-parameter, $b$, space ($b = b_{\perp}$ is Fourier conjugate
to $q'_{\perp}$):
\begin{equation}
A(s,t,r^2) \equiv 2 s\int d^2b\, e^{i \vec{q^{'}} \cdot \vec{b}} f(s,b,r^2) \ .
\label{eq:nuc:black_nuc:bbl:1}
\end{equation}
The scattering amplitude $f(s,b,\dperp^2)$ is related to the
total cross section for the scattering
of dipoles of fixed transverse diameter, $\dperp$, by the optical theorem:
\begin{equation}
\sigma_{\mbox{{\footnotesize tot}}} (s,\dperp^2)=2 \int d^2 b \,Im\, f(s,b,\dperp^2) \ .
\label{eq:nuc:black_nuc:bbl:2}
\end{equation}
The unitarity of the scattering $S$-matrix \cite{eden}
($S_{ab} =  \delta_{ab} + i f_{ab}, S^{\dagger}_{ac} S_{cb} = \delta_{ab}$) imposes
the following conditions \footnote{A rather straightforward way
to deduce unitarity of
the $S$-matrix is to consider amplitudes for the scattering of
bound states of fictitious
heavy quarks $Q$ and the prove that, for sufficiently large
mass of the quarks $M_Q$ and
small $x$, dipole scattering would dominate. In order to
complete the derivation,
one also needs to use the fact that QCD is flavour blind
if the resolution scales are chosen appropriately.}
on the scattering amplitudes (for each $b$, a continuum
analogy of angular momentum in a partial wave analysis in 
non-relativistic quantum mechanics):
\begin{equation}
Im f_{aa}(s,b,\dperp^2) =\frac{1}{2}( f^{\dagger}_{aa} f_{aa} +
\Sigma_{c, c\neq a} f^{\dagger}_{ac} f_{ca} ) \, ,
\label{eq:nuc:black_nuc:bbl:3}
\end{equation}
\noindent where ``a,c'' are labels for definite states (on the right hand side 
we suppress the variables $s,b,\dperp$ for clarity).
The first and second terms on the right hand side of (\ref{eq:nuc:black_nuc:bbl:3}) 
involve elastic and inelastic final states, respectively.

The black body limit (BBL) assumes that: (i) configurations with the impact parameters
satisfying $b^2 \leq b^2_{\mbox{{\footnotesize max}}}$
are completely absorbed by target, i.e. the elastic matrix elements of the 
$S$-matrix are zero for those impact parameters: 
$S_{aa}(b \leq b_{\mbox{{\footnotesize max}}})=\delta_{aa}+if_{aa} (b \leq b_{\mbox{{\footnotesize max}}})=0$.
 This implies that $Im f_{aa}(s,b<_{\mbox{{\footnotesize max}}},\dperp^2)=1$,
and leads to $|f_{aa}(s,b,\dperp^2)|^2=\Sigma_{c, c\neq a} |f_{ac}(s,b,\dperp^2)|^2$ 
that is to the equality of the elastic and inelastic contributions to the total 
cross section.
(ii) the region $b^2 \leq b^2_{\mbox{{\footnotesize max}}}$ gives the dominant contribution to the
scattering amplitude (\ref{eq:nuc:black_nuc:bbl:2}). The great advantage
of the BBL is that calculations of the amplitudes of some small-$x$ processes
do not require specific model assumptions. Moreover, as will be
discussed later,
the BBL approximation seems to be realistic for DIS on heavy nuclear targets.

The dependence of $f(s,b,\dperp^2)$ on the impact parameter, $b$, at large $b$
follows from analytic properties of the scattering amplitude
$A(s,t,\dperp^2)$ in $t$-plane.
A simple analysis leads to $f(s,b,\dperp^2)=c\, \exp( -\mu b)$ at large $b$,
where $c \propto xG^T(x,\dperp^2) \propto 1/x^{n}$. 
Hence, following Froissart \cite{Froissart,eden}
we may evaluate the maximal impact parameter characterising the black
body limit.
One obtains $b^2_{\mbox{{\footnotesize max}}} \propto 1/\mu^2 \ln^{2} 1/x$
for a nucleon target, and
$b^2_{\mbox{{\footnotesize max}}} = R_A^2$ for a heavy nuclear target
($R_A$ being the radius of the nucleus). Note that the
difference between $b^2_{\mbox{{\footnotesize max}}}$ for the nucleon and
nuclear case reflects the fact that the nucleon target is not a homogeneous
sphere but rather an object with an extended diffuse edge.

Using these relationships, for the total dipole-nucleon scattering
cross section (see also (\ref{eq:nuc:black_nuc:bbl:2})) in the BBL approximation,
one obtains
\begin{equation}
{\hat \sigma}_{\mbox{{\footnotesize tot}}} (s,\dperp^2) = 2\pi (R_N^2 + 4c_N \ln^2 1/x) \, .
\label{eq:nuc:black_nuc:bbl:4}
\end{equation}
In addition to the total cross section, one can
examine the $t$-dependence of the
cross section, corresponding to the amplitude in 
(\ref{eq:nuc:black_nuc:bbl:1}), defined
through its slope, $B$:
\begin{equation}
\frac{d \ln \sigma(s,t)}{dt}\Big|_{{t=0}}=B=B_0+2\alpha_{\mbox{{\footnotesize eff}}}^{\prime}\ln 1/x =
\frac{\int d^2 b\, b^2 f(s,b,\dperp^2)}{2 \int d^2 b\, f(s,b,\dperp^2)} \ ,
\label{eq:nuc:black_nuc:bbl:5}
\end{equation}
\noindent where $\alpha_{\mbox{{\footnotesize eff}}}^{\prime} \equiv d~(B/2) / d \ln 1/x$.
We use above that the $t$-dependence of the real and imaginary parts of
the amplitude is the same.
In the BBL, one obtains
\begin{eqnarray}
&&B=\frac{b^2_{\mbox{{\footnotesize max}}}}{4} =\frac{R_T^2}{4}+c_T \ln^2 1/x 
\, , \nonumber \\
&&\alpha_{\mbox{{\footnotesize eff}}}^{\prime} = c_T \ln 1/x  \ .
\label{eq:nuc:black_nuc:bbl:6}
\end{eqnarray}
Here $R_T$ is the radius of the target, $c_T$ is a 
factor which is similar for nucleon and nucleus targets.
Hence for practical purposes, one can 
neglect $c_T$ for heavy nuclear targets.

Finally, the proton structure function (at fixed $Q^2$) in the BBL reads
\begin{eqnarray}
F^{p}_{2}(x,Q^2) \propto \sum_{i} e_i^2 Q^2 {2\pi R_N^2\over 12 \pi^3}
(1+{4c_N^2\over R_N^2} \ln^2 x_0/x) \ln 1/x \, .
\label{eq:nuc:black_nuc:bbl:6p}
\end{eqnarray}
Here the sum is taken over electric charges $e_i$ of active quark flavours $i$.
Note that the additional factor of $\ln 1/x$ in 
(\ref{eq:nuc:black_nuc:bbl:6p}) as compared
to (\ref{eq:nuc:black_nuc:bbl:4}) is due to the
contribution of large masses resulting from the singular nature of the photon 
wavefunction. 
This reflects the logarithmic divergence of renormalization coupling constant
for the electric charge (cf. (\ref{eq:nuc:black_nuc:bbl:8}) below).
As one sees from (\ref{eq:nuc:black_nuc:bbl:6p}), general principles of QCD
and, in particular, the BBL approximation, do not exclude a
fast ($\ln^3 1/x$) increase of the structure functions of a nucleon 
at $x\to 0$.
In addition, the contribution from dipoles with the impact parameters larger
than $b_{\mbox{{\footnotesize max}}}$ or sufficiently small dipoles
(which are assumed to give a small contribution in the BBL approximation)
should continue to increase with increasing energy as dictated by the
DGLAP approximation of QCD.
Thus, in practice, it would be very difficult, or even impossible,
to distinguish the BBL prediction (\ref{eq:nuc:black_nuc:bbl:6p})
from a similarly rapid growth predicted by the DGLAP equation and
to search for saturation effects in inclusive nucleon structure
functions. Hence, one should turn to DIS on nuclear targets in order 
to search for distinct signals of BBL dynamics.

The use of nuclei has two clear advantages. Firstly, scattering
at large impact parameters,
where the interaction is far from the BBL over a wide range of energies (the
edge effects), is suppressed by the factor $R_N/R_A$. Secondly, in
a broad range of
impact parameters, $b \leq R_A$, the nuclear thickness is practically
$b$-independent and much larger than in a nucleon. Hence, DIS on sufficiently
heavy nuclei can serve as a good testing ground for the application of
the BBL and will allow
us to pinpoint some distinctive features of it. For example, as follows
from the above discussion, the
unitarity of $S$-matrix significantly tames the rapid grows of the
nuclear structure functions and predicts the unitarity limit
\begin{equation}
F^{A}_{2}(x,Q^2) = \sum_{i} {e_i^2} Q^2 {2\pi R_A^2\over 12 \pi^3} 
\ln 1/4m_NR_Ax \ .
\label{eq:nuc:black_nuc:bbl:7}
\end{equation}

Over the last few years a number of models, using the
infinite momentum frame, were suggested in order to explain the dynamics of
DIS at small $x$ by building the nuclear wave function from
large gluon fields and
assuming a certain saturation of the parton densities
(for the review and references see
e.g., \cite{Mueller}). In many respects, these models and the
BBL approximation are similar.

The BBL in DIS from a heavy nucleus at small $x$ was first considered
by Gribov \cite{gribov}, before the discovery of QCD.
Gribov assumed that each hadronic fluctuation of
the virtual photon interacts with the target nucleus with
the same maximal strength allowed by unitarity. Such an
assumption, supported by the observed cross sections of
hadron-nucleus interactions, was natural for understanding the dynamics of
soft strong interactions in models predating QCD.

Thus, in the black body limit, DIS on nuclei is dominated by the dissociation
of the incoming virtual photon into its hadronic fluctuations,
which subsequently
interact with the target with the same scattering cross sections $2\pi R_A^2$,
and then hadronize into final states with mass $M$.
The transverse and longitudinal nuclear structure functions 
may be conveniently formulated as an integral over produced masses
\begin{eqnarray}
&&F^A_T (x,Q^2)= C \int_{M^2_{\mbox{{\footnotesize min}}}}^{
M^{2}_{\mbox{{\footnotesize max}}}} dM^2 ~\frac{2\pi R_A^2}{12 \pi^3}
\frac{Q^2 M^2 \rho (M^2)}{(M^2+Q^2)^2} \, , \\
&&F^A_L (x,Q^2)= C \int_{M^2_{\mbox{{\footnotesize min}}}}^{
M^{2}_{\mbox{{\footnotesize max}}}} dM^2 ~\frac{2\pi R_A^2}{12 \pi^3}
\frac{Q^4     \rho (M^2)}{(M^2+Q^2)^2} \, , 
\label{eq:nuc:black_nuc:bbl:8}
\end{eqnarray}
where $\rho (M^2)=\sigma^{e^ + e^- \rightarrow \mbox{{\footnotesize hadrons}}} (M^2)/\sigma^{e^+e^-\to \mu^+\mu^-} (M^2)$. In the BBL, the
coefficient $C = 1$. The upper cutoff, $M^2_{\mbox{{\footnotesize max}}} \ll W^2 \approx 2q_0 m_{N}$, comes from the nuclear form factor:
\begin{equation}
-\frac{t_{\mbox{{\footnotesize min}}} R_A^2}{3} \approx \frac{(M^2+Q^2)^2}{4q_0^2} R_A^2/3 \approx
 m_{N}^2 x^2 \frac{R_A^2}{3} \ll 1 \, .
\label{eq:nuc:black_nuc:bbl:9}
\end{equation}
The key element of the derivation of (\ref{eq:nuc:black_nuc:bbl:8})
is the observation that in the BBL, as a result of orthogonality
of the wave functions of the
eigenstates of QCD Hamiltonian with different energies,
the non-diagonal transitions between states with different $M^2$
are zero. This enables one to write the structure functions as
a single dispersive integral
as is done in (\ref{eq:nuc:black_nuc:bbl:8}). Since
(\ref{eq:nuc:black_nuc:bbl:8})
leads to a cross section for $\gamma^{\ast}A$ scattering grossly
violating Bjorken scaling
($\sigma_{\mbox{{\footnotesize tot}}}^{\gamma^{\ast}A}(x,Q^2) \propto \pi R_A^2 \ln(1/x)$
instead of $\propto 1/Q^2$),
the BBL has been considered for some time to be an artifact of
the pre-QCD physics.
This is especially so since, within the parton model, the aligned
jet model removed this gross scaling violation \cite{Bjorken}. However,
for consistency of the
target rest frame and infinite momentum frame descriptions, an
exponential suppression with
decreasing $\dperp^2$ of the cross section of interaction of
small-size configurations with
hadrons was assumed to be required. This fact was subsequently
explained and understood in terms of
the QCD factorization theorem for the scattering of small
dipoles, colour neutrality of the dipole and
asymptotic freedom for hard processes in QCD (and the
``suppression'' was realised only to be only a single
power in $\dperp^2$, cf. (\ref{eq:nuc:intr:eqshat})).

In perturbative QCD, the dipole-target cross
section (\ref{eq:nuc:intr:eqshat}), rapidly
increases with increasing energy since the gluon
density rapidly increases with decreasing $x$.
Hence, if the increase of the interaction cross section is
not tamed by some mechanism,
it will reach values expected for the black body limit
(i.e. tens of mb, cf. (\ref{eq:nuc:intr:sigblack})).

Properties of the BBL in QCD are somewhat different from
those within the Gribov picture due to
a significant probability of smaller than average size configurations in the
photon wave function, for which the conditions of the black
body limit are not satisfied.
As a result, the interaction of such small-size configurations is not tamed.
Thus, in contrast to the Gribov approach, only a fraction of
all configurations
will interact according to the BBL approximation and therefore $C < 1$ in
(\ref{eq:nuc:black_nuc:bbl:8}).

Using (\ref{eq:nuc:intr:eqshat}), it is straightforward to
estimate the kinematical boundaries where the unitarity limits may be reached.
Indeed, the requirement that $\sigma_{\mbox{{\footnotesize el}}} \leq \sigma_{\mbox{{\footnotesize tot}}}/2$ \cite{fks1,afs}
(also assuming that (\ref{eq:nuc:intr:eqshat}) is applicable for
the range of
$x$ in which the taming is necessary) indicates that
for some gluon-induced hard processes
the unitarity limit should be well within the reach of an electron-nucleus
collider at HERA/THERA (for a review see \cite{FGSreview}).

One can also make an interesting prediction about nuclear shadowing.
Since the $x$-dependence of the nuclear structure function of
(\ref{eq:nuc:black_nuc:bbl:7})
is significantly weaker than that for the proton structure function
of (\ref{eq:nuc:black_nuc:bbl:6p}),
nuclear shadowing is not saturated (as is often assumed in the
limit of fixed $Q^2$ and $x\to 0$), and we find
\begin{equation}
\frac{F^{A}_{2}(x,Q^2)}{A F^{N}_{2}(x,Q^2)}\propto {R_A^2\over A R_N^2}
{1\over 1+{4c_N^2 R_A^2} \ln^2 1/x} \ .
\label{eq:nuc:black_nuc:bbl:12}
\end{equation}
However, at unrealistically small $x$, where the impact parameters
become significantly
larger than $R_A$, this effect will disappear.

In summary, since the contributions of small configurations
(which have not reached the black body limit)
remain significant in a wide range of $x$ and $Q^2$,
studies of the total cross sections are a rather ineffective way to the
search for the onset of the BBL regime. In particular,
it may be rather difficult to
distinguish the BBL from DGLAP approximation with different
initial conditions in this way.

In what follows we shall demonstrate that studies of
DIS final states provide a number of
clear signatures of the onset of the BBL regime, which will be
qualitatively different from the leading
twist regime. For simplicity, we will assume that the BBL is
reached for a significant
part of the cross section and, hence, restrict our discussion to
DIS on a large nucleus so that
edge effects (which are important in the case of scattering off a nucleon)
can be neglected.

\subsection{Diffractive final states}

\label{sec:nuc:black_nuc:dfs}

The use of Gribov's orthogonality argument (to neglect
non-diagonal transitions
in (\ref{eq:nuc:black_nuc:bbl:8})) allows the integrals over the masses
to be removed in the expressions for the structure functions and, since
diffraction is 50\% of the total cross section in the BBL, we
immediately find for the spectrum of diffractive masses:
\begin{eqnarray}
&&{dF_T^{D(3)}(x,Q^2,M^2)\over dM^2}=
{\pi R_A^2\over 12 \pi^3}
{Q^2M^2\rho(M^2)\over (M^2+Q^2)^2} \ , \nonumber\\
&&{dF_L^{D(3)}(x,Q^2,M^2)\over dM^2}=
{\pi R_A^2\over 12 \pi^3}
{Q^4\rho(M^2)\over (M^2+Q^2)^2} \ .
\label{eq:nuc:black_nuc:dfs:1}
\end{eqnarray}

\noindent Moreover, the spectrum of hadrons in the centre of mass of
the diffractively produced system should be the same as
in $e^+e^-$ annihilation.
Hence, the dominant diffractively-produced final state will have two jets
with a distribution over the centre of mass emission angle proportional to
$1+\cos^2 \theta$ for the transverse case and $\sin^2 \theta$ for the
longitudinal case:
\begin{equation}
{dF_T^{D(3)}(x,Q^2,M^2)\over dM^2 d\cos \theta}=
\frac{3}{8} (1 + \cos^2 \theta) {\pi R_A^2\over 12 \pi^3}
{Q^2M^2\rho(M^2)\over (M^2+Q^2)^2} \ ,
\label{eq:nuc:black_nuc:dfs:2}
\end{equation}
\begin{equation}
{dF_L^{D(3)}(x,Q^2,M^2)\over dM^2d\cos \theta}=\frac{3}{4}
\sin^2 \theta {\pi R_A^2\over 12 \pi^3} {Q^4\rho(M^2)\over (M^2+Q^2)^2} \ .
\label{eq:nuc:black_nuc:dfs:3}
\end{equation}

The transverse momentum of the jet, $p_{t}$, and the longitudinal fraction
of photon energy, $z$, carried by the jet are related to the
diffractive mass, $M$, and the angle, $\theta$, as follows
(we neglect here the quark masses as compared to $Q, M$):
\begin{eqnarray}
&&p_t={M\over 2}\sin \theta \ , \nonumber\\
&&z=(1+\cos \theta)/2 \ .
\label{eq:nuc:black_nuc:dfs:4}
\end{eqnarray}
Hence, in the BBL, diffractive production of high
$p_t$ jets is strongly enhanced:
\begin{eqnarray}
&&\left<p_t^2(jet)\right>_T
=3M^2/20 \ , \nonumber\\
&& \left<p_t^2(jet)\right>_L
=M^2/5 \ .
\label{eq:nuc:black_nuc:dfs:5}
\end{eqnarray}
This is to be compared to the leading twist approximation
where it is $\propto \ln Q^2$.
The relative rate and distribution of the jet variables for the three
jet events (originating from $q\bar q g $ configurations) will be also
the same as in $e^+e^-$ annihilation and hence is given by the standard
expressions for the process $e^+e^- \to q\bar q g$ (see e.g. \cite{book}).

An important advantage of the diffractive BBL signal is that
these features of the diffractive final state should hold for
$M^2 \leq Q^2_{BBL}$ even for $Q^2 \geq Q^2_{BBL}$
because configurations with transverse momenta
$\leq Q_{BBL}/2$ still interact in the black regime
(and correspond to transverse size fluctuations for which the
interaction is already black).

Another interesting feature of the BBL is the spectrum of the leading
hadrons in the virtual photon fragmentation region produced in DIS.
The spectrum is essentially given by
the $\theta$-dependence of
(\ref{eq:nuc:black_nuc:dfs:2}) and (\ref{eq:nuc:black_nuc:dfs:3}).
For fixed $M^2$,
the jet distribution over $z$ for transversely polarised photons is simply
\begin{equation}
{d\sigma_T \over d z}\propto 1+(2z-1)^2 \, .
\label{eq:nuc:black_nuc:dfs:6}
\end{equation}
Similarly, for longitudinally polarised photons,
\begin{equation}
{d\sigma_L \over d z}\propto ~z(1-z) \, .
\label{eq:nuc:black_nuc:dfs:7}
\end{equation}
\noindent If no special separation procedure is undertaken, at small $x$ one
actually measures $\sigma_L+\epsilon\sigma_T$
($\epsilon$ is the photon polarisation). In this case,
combing (\ref{eq:nuc:black_nuc:dfs:2}) and
(\ref{eq:nuc:black_nuc:dfs:3}) we find:
\begin{equation}
{d(\sigma_T + \epsilon \sigma_L)\over d z}\propto {M^2\over 8 Q^2}
(1+(2z-1)^2) + \epsilon z(1-z) \, .
\end{equation}

Exclusive vector meson production in the BBL is in a sense a resurrection of
the original vector meson dominance model \cite{Sakurai} without off-diagonal
transitions. The amplitude for the vector meson-nucleus interaction
is proportion
to $2\pi R_A^2$ (since each configuration in
the virtual photon interacts with the same BBL cross section).
This is markedly different from the requirements \cite{FGSvdm} for matching
generalised vector dominance model (see e.g., \cite{VDM})
with QCD in the scaling limit, where the non-diagonal matrix elements
are large and lead to strong cancellations.
Hence, we can factorize out the cross section for the dipole interaction 
from the overlap integral between wavefunctions of virtual photon and vector meson
to obtain for the dominant electroproduction of vector mesons
\begin{eqnarray} &&
{d\sigma^{\gamma^{\ast}_{T} +A\to V+A} \over dt} 
 =
{M_V^2\over Q^2} {d\sigma^{\gamma^{\ast}_{L} +A\to V+A} \over dt}= 
\nonumber \\ &&
{(2\pi R_A^2)^2\over 16\pi}{3 \Gamma_V M_V^3 \over \alpha
(M_V^2 + Q^2)^2 } \frac{4~\left|J_1(\sqrt{-t}R_A)\right|^2}{-tR_A^2},
\label{vm}
\end{eqnarray}
where $\Gamma_V$ is the electronic decay width $V\to e^+e^-$,
$\alpha$ is the fine-structure constant, and $J_1(x)$ 
is the Bessel function.
Thus the parameter-free prediction is that in the BBL
at large $Q^2$ vector meson production
cross sections have a $1/Q^2$ behaviour, in stark contrast to
an asymptotic behaviour of
$1/Q^6$ predicted in perturbative QCD \cite{brod,fks1,fks2}.

In order to observe the onset of the BBL regime for a nucleon target, 
one should
consider scattering at small impact parameters.
However, a direct comparison of the BBL prediction for small $b$ with data is very difficult
since, in the BBL, the amplitude oscillates as a function of $t$. Also, in the
case of a proton target, a significant nucleon spin-flip amplitude may mask
these oscillations.

\subsection{Inclusive spectra}

\label{sec:nuc:black_nuc:is}

In the leading twist approximation, the QCD factorization theorem is valid and
leads to universal spectra of leading particles (independent of the target)
for the scattering off partons of the same flavour. Fundamentally,
this can be
explained by the fact that, in the Breit frame, the
fast parton which is hit by
the photon carries practically all of the photon's
light-cone momentum ($z \to 1$).
Due to QCD evolution, this parton acquires virtuality
$\sim Q^2$, and a rather large
transverse momentum, $k_t$ (which is still $\ll Q^2$). So,
in pQCD, quark and gluons
emitted in the process of QCD evolution and in the
fragmentation of highly virtual
partons together still carry all the photon momentum.
In contrast, in the BBL, configurations in which partons carry
all of the photon momentum
form only part of cross section of leading hadron production.
Another part is from inelastic collisions of configurations where all
partons carry appreciable momentum fractions and large relative
transverse momenta
(these configurations are rather similar to the case of diffractive scattering
(see e.g. (\ref{eq:nuc:black_nuc:dfs:2}) and (\ref{eq:nuc:black_nuc:dfs:3}))).
Hence, in the BBL case, the spectrum of leading hadrons (in
the direction of the virtual photon) is expected to be much depleted.

The inclusive spectrum of leading hadrons 
can be estimated as due to the  
independent fragmentation of quark and antiquark of
 virtualities $\geq Q^2$, with $z$ and  $p_{\perp}$ distributions
given by 
(\ref{eq:nuc:black_nuc:dfs:2}) and
(\ref{eq:nuc:black_nuc:dfs:3}) (cf. the case of
diffractive production of jets discussed above).
The independence of fragmentation is justified because large transverse momenta of quarks 
dominate in the photon wave function  (cf. eqs.~(\ref{eq:nuc:black_nuc:dfs:2}, \ref{eq:nuc:black_nuc:dfs:3}, \ref{eq:nuc:black_nuc:dfs:4})
and because of the weakness of the final state interaction between $q$ and $\bar q$, since the
$\alpha_s$ is small and the rapidity interval is of the order of one.
Obviously, this leads to a gross depletion of the
leading hadron spectrum as compared to the leading twist
approximation situation where leading
hadrons are produced in the fragmentation region of the parton which carries
essentially all momentum of the virtual
photon \footnote{Qualitatively, this pattern is similar
to the one expected in
the soft region since the spectrum of hadrons produced in hadron-nucleus
interactions is much softer than for hadron-nucleon interactions.}.
Taking the production of multi-jet states like $q\bar q g$ into account will
further enhance this scaling violation.
If we neglect gluon emissions in the photon wave function, we find,
for instance:
\begin{equation}
{d N^{\gamma_{T}^{\ast}/h}\over
dz}=
2~\int_z^1 D^{q/h}(z/y,Q^2){3\over 4}(1+(2y-1)^2)dy \, ,
\label{eq:nuc:black_nuc:dfs:41}
\end{equation}
\noindent for the total differential 
multiplicity of leading hadrons produced by
transversely polarised virtual photons, in the BBL.
$D^{q/h}(z/y,Q^2)$ is the fragmentation function of a quark with 
flavour $q$ into the hadron $h$.
Here we use that $D^{u/h}(z/y,Q^2)=D^{d/h}(z/y,Q^2)$
and neglect a small difference in the fragmentation functions of 
light and heavy quarks.

\begin{figure}
\begin{center}
\epsfig{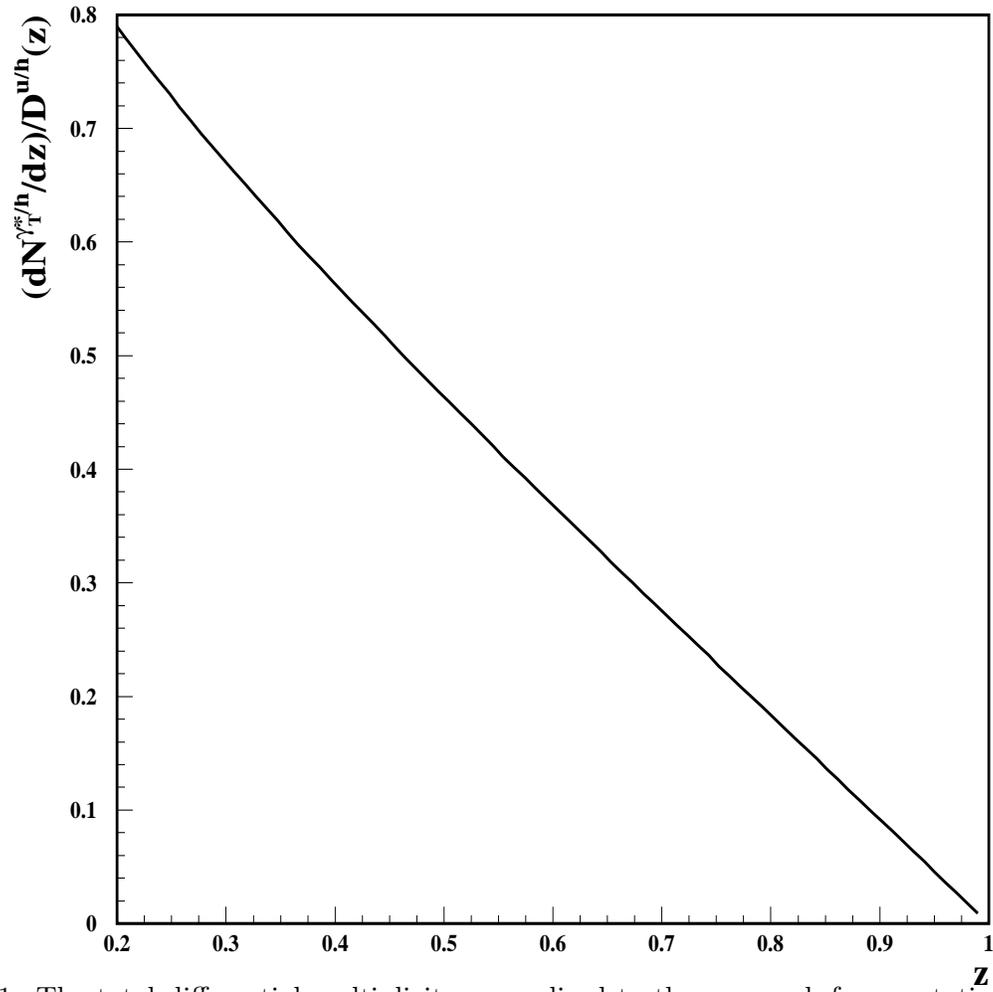}
\caption{The total differential multiplicity normalized to the up quark fragmentation function \protect\cite{Bourhis}, $(d N^{\gamma_{T}^{\ast}/h} /dz)/D^{u/h}(z,Q^2)$, 
as a function of $z$ at $Q^2$=2 GeV$^2$ calculated in the BBL using (\ref{eq:nuc:black_nuc:dfs:41}).}
\label{fig:nuc:black_nuc:1}
\end{center}
\end{figure}

An illustration of the results of the calculation
of $d N^{\gamma_{T}^{\ast}/h} /dz$ is presented in Fig.~\ref{fig:nuc:black_nuc:1}. 
We normalise the distribution to the leading twist case by using realistic 
up-quark fragmentation functions at $Q^2 =2$~GeV$^2$ \cite{Bourhis} 
(the up and down quark distributions are similar and we neglect the small
difference induced by including further flavours). One can see
from the figure that a gross violation of the
factorization theorem is expected in the BBL.
The spectrum of leading hadrons is much softer
at large $z$, with an excess multiplicity
at $z\le 0.1$. Note that the use of leading twist fragmentation functions in
the above expression probably underestimates absorption.
So the  curve in Fig.\ref{fig:nuc:black_nuc:1} can
be considered as a conservative lower limit for the amount of suppression.

With an increase of $Q^2$ we expect a further softening related to
a change in the partonic structure of the virtual photon wave function.
Progressively more configurations contain extra hard gluons, which fragment
independently in the BBL, further amplifying deviations from the
standard leading twist predictions.

Another important signature of the BBL is the change of
$p_t$ distributions with
decreasing $x$ (at fixed $Q^2$). The spectrum
of the leading hadrons should broaden due to increased $p_t$ of the
fragmenting partons. Hence, the most efficient strategy would be to
select leading jets in the current fragmentation region and examine the
$z$ and $p_t$-dependence of such jets. Qualitatively, the effect of
broadening of $p_t$-distributions is similar to the increase of the $p_t$
distribution in the model \cite{Mclerran}, although final states
in DIS were not
discussed in this model.

An important advantage of inclusive scattering off
a nucleus is the possibility
to use a centrality trigger. For example, one could use the number of nucleons
emitted in the nucleus decay (soft nucleons in the nucleus rest frame)
to select scattering at the central impact parameters. Such a
selection allows the
effective thickness of the nucleus to be increased, as compared
to the inclusive situation,
by a factor $\sim 1.5$ and, hence, allows for the BBL to be
reached at significantly
larger $x$.
The signal for the BBL will be a change of the spectrum with centrality of the collisions, in
contrast to the LT case where no such correlation is expected. Note that the lack of absorption 
of leading particles in DIS off nuclei at the fixed target energies and $Q^2 \ge 2$~GeV$^2$ is well
established experimentally, see e.g.  \cite{Ashman}.

To summarise, predictions for a number of simple final state
observables in the
black body limit are distinctly different from those made in the leading twist
approximation and, hence, will provide model-independent tests of
the onset of the BBL.

\section{Unitarity constraints for electron-nucleus DIS}
\label{sec:nuc:unit}

While predictions for inclusive DIS and DIS final state
observables in the BBL are
distinct, one still would like to determine the kinematical
region where the regime of
the BBL sets in. One way to address this issue it to examine
the unitarity of the $S$-matrix
for the interactions of purely perturbative QCD fluctuations.

\subsection{The interaction of small colour
dipoles with hadrons}
\label{sec:nuc:unit:sm}

The dipole picture of electron-target DIS is valid
when the lifetime of the fluctuation is
much longer than the interaction time with the target. Equally, this relation
may be expressed in terms of the {\it coherence length},
$l_{coh}$, of the fluctuation
(relative to the target radius). This length is given by
the average longitudinal distances (Ioffe distances) in the correlator of the
electromagnetic currents which determine the structure function $F_{2}(x,Q^2)$.
A simple analysis shows that $l_{coh} \sim 1 / 
(2 m_N x)$. At HERA $l_{coh}$ can reach values of $10^3$~fm (for moderate $Q^2$ only).
As $Q^2$ increases, the scaling violations lead to a gradual
reduction of $l_{coh}$ at fixed $x$
and ultimately to the dominance of longitudinal distances
$\sim R_N$ (for a discussion see \cite{FGSreview}).

In the target rest frame, when $l_{coh}$ is sufficiently large,
the virtual photon
can fluctuate into a variety of partonic configurations containing
various numbers of partons and involving different transverse sizes. The
simplest is a $q \bar{q}$ dipole, which dominates at very small distances.
Since under SU(3)$_{c}$ transformations the quark and the anti-quark
transform in the fundamental representation
(like 3 and ${\bar 3}$) the name ``colour triplet'' is often used
for the $q \bar{q}$ pair (although overall its colour is of course neutral).
The $q \bar{q}$-dipole of a small diameter interacts with a
hadronic target with an
interaction cross section given by (\ref{eq:nuc:intr:eqshat}).
Another important photon fluctuation is the one consisting of
a quark, anti-quark
and gluon, with a relatively large transverse momentum between
the quark and the anti-quark.
Such configurations effectively transform in the
adjoint representation (effectively $8$,
${\bar 8}$) and so are known as the $q \bar{q}g$ {\it colour octet dipole}.
There are of course other $q {\bar q} g $ configurations,
for example those in which
the gluon and either the quark or anti-quark have a large
relative transverse momentum
(these merely correspond to the ${\cal O} (\alpha_s)$
corrections to the colour triplet dipole)
and other more general configurations which do not
correspond to dipoles at all.

For the case of scattering of colour octet dipoles off a
target, the corresponding cross
section is enhanced a colour factor (given by the ratio of
the Casimir operators of
SU(3)$_c$), $C_F(8)/C_F(3)= 9/4$ \cite{afs,rad,bhm}:
\begin{eqnarray}
{\hat \sigma}^{{\small \mbox{{\footnotesize octet}}}}_{\mbox{{\scriptsize pQCD}}} (\dperp^2,x) & = &
{\Large{{9 \over 4}}} \, {\hat \sigma}^{{\small \mbox{{\footnotesize triplet}}}}_{\mbox{{\scriptsize pQCD}}} (\dperp^2,x) \nonumber \\
& = &\frac{3 \pi^2}{4} \, \dperp^2 \, \alpha_s ({\bar Q^2}) \, x G_T(x,{\bar Q^2}) \, .
\label{eq:nuc:unit:sm:1}
\end{eqnarray}
\noindent Both (\ref{eq:nuc:intr:eqshat}) and (\ref{eq:nuc:unit:sm:1})
predict cross sections steeply rising with increasing energy (driven
by the rise of
the gluon density with decreasing $x$).
Fitting the energy dependence in the form
\footnote{Such fit is useful in practical applications.
Perturbative QCD predicts the behaviour $\propto a +b\ln 1/ x
+ c\ln^2 1 /x$ for the HERA energy range
since radiation of $\leq$ 1-2 hard gluons is possible in the multi Regge
kinematics at HERA.}
$\sigma_{\mbox{{\scriptsize pQCD}}} (s,Q^2)\propto s^{n(Q^2)}$, one finds
\begin{eqnarray}
&&n~(Q^2=~4~\mbox{GeV}^2) \approx 0.2 \ , \nonumber\\
&&n~(Q^2=40~\mbox{GeV}^2)\approx 0.4 \ .
\label{eq:nuc:unit:sm:2}
\end{eqnarray}
One of the manifestations of the behaviour predicted
by (\ref{eq:nuc:unit:sm:2}) is the $Q^2$-behaviour of the
measured exclusive vector meson production. The interaction cross sections of
(\ref{eq:nuc:intr:eqshat}) and (\ref{eq:nuc:unit:sm:1})
may be thought of as a
complimentary
description of the physics described by the leading log
QCD evolution equations at small $x$.
However, an accurate determination of the relation between ${\bar Q^2}$
and the transverse size $\dperp$ in these equations
requires a next-to-leading order QCD analysis in this
framework which has not been done yet.
Numerical studies \cite{fks2,mfgs} based on matching of
the $\dperp$-space and $Q$-space expressions
for $\sigma_L(x,Q^2)$, lead to $\lambda \sim 9-10 $ for a
sufficiently broad range of $Q^2$ and $x$. With this choice of
$\lambda$, a good description of the recent inclusive
DIS electron-proton data was obtained (with a suitable
extrapolation to large $\dperp$ \cite{mfgs})
without any further fitting. As already mentioned,
$\lambda$ is a logarithmically-decreasing
function of the dipole size $\dperp$. In particular, at large
values of $\dperp$, $\dperp \ge 0.3$ fm, where connection between
$\dperp^2$ and $Q^2$ is rather sensitive to non-perturbative
effects, one expects a decrease of $\lambda$ with increasing
$\dperp^2$. On the other hand, it was found that
variations in $\lambda$ do not significantly effect values
of $\hat{\sigma}_{\mbox{{\scriptsize pQCD}}} (\dperp^2,x)$.
Hence, our following estimates of the unitarity constraints,
which are made using
(\ref{eq:nuc:intr:eqshat}) and (\ref{eq:nuc:unit:sm:1}), are
insensitive to a possible decrease of $\lambda$ at large $\dperp^2$.

\subsection{Unitarity constraints}
\label{sec:nuc:unit:uc}

The rapid increase of the cross sections given in
(\ref{eq:nuc:intr:eqshat}) and (\ref{eq:nuc:unit:sm:1}) with
decreasing $x$ cannot continue forever, otherwise the
unitarity of the $S$-matrix will be
violated (see (\ref{eq:nuc:black_nuc:bbl:3})). The unitarity
boundary $Im~f(s,b,\dperp^2)=1$ can be expressed in
terms of the pQCD dipole cross section as
\begin{equation}
{\hat \sigma}_{\mbox{{\scriptsize pQCD}}}^{\mbox{{\footnotesize inel}}}=\sigma^{\mbox{{\footnotesize el}}}=\sigma^{\mbox{{\footnotesize tot}}}/2 \ ,
\label{eq:nuc:unit:uc:1}
\end{equation}
\noindent and is applicable to any hadronic target.
For a nucleus, with the atomic number
$A$, ${\hat \sigma}_{\mbox{{\scriptsize pQCD}}}^{\mbox{{\footnotesize inel}}}$ cannot exceed its geometric limit
$\pi R^2_{A}$. Thus, generalising
(\ref{eq:nuc:intr:eqshat}) and (\ref{eq:nuc:unit:sm:1}) for a
nuclear target, we obtain
the following kinematical restrictions imposed by the unitarity
of the $S$-matrix for $x \ll 1 / 4 R_A m_N$:
\begin{eqnarray}
&&\sigma^{q \bar{q}}_{A}(\dperp^2,x)
= \frac{\pi^2}{3} \dperp^2 \left[ x
G^{A}(x, {\bar Q^2}) \right] \alpha_{s}({\bar Q^2})
\lsim \pi R_A^2 , \nonumber\\
&&\sigma^{q \bar{q}g}_{A}(\dperp^2,x)
= \frac{3\pi^2}{4} \dperp^2 \left[ x
G^{A}(x,{\bar Q^2}) \right] \alpha_{s}({\bar Q^2})
\lsim \pi R_A^2\ ,
\label{eq:nuc:unit:uc:2}
\end{eqnarray}
where $xG^{A}(x,{\bar Q^2})$ is the nuclear gluon density.
The kinematical boundaries
following from these equations are presented as
curves in the $x$-${\bar Q}$ plane in Figs.~14-17 of \cite{FGSreview}.

\begin{figure}
\begin{center}
\epsfig{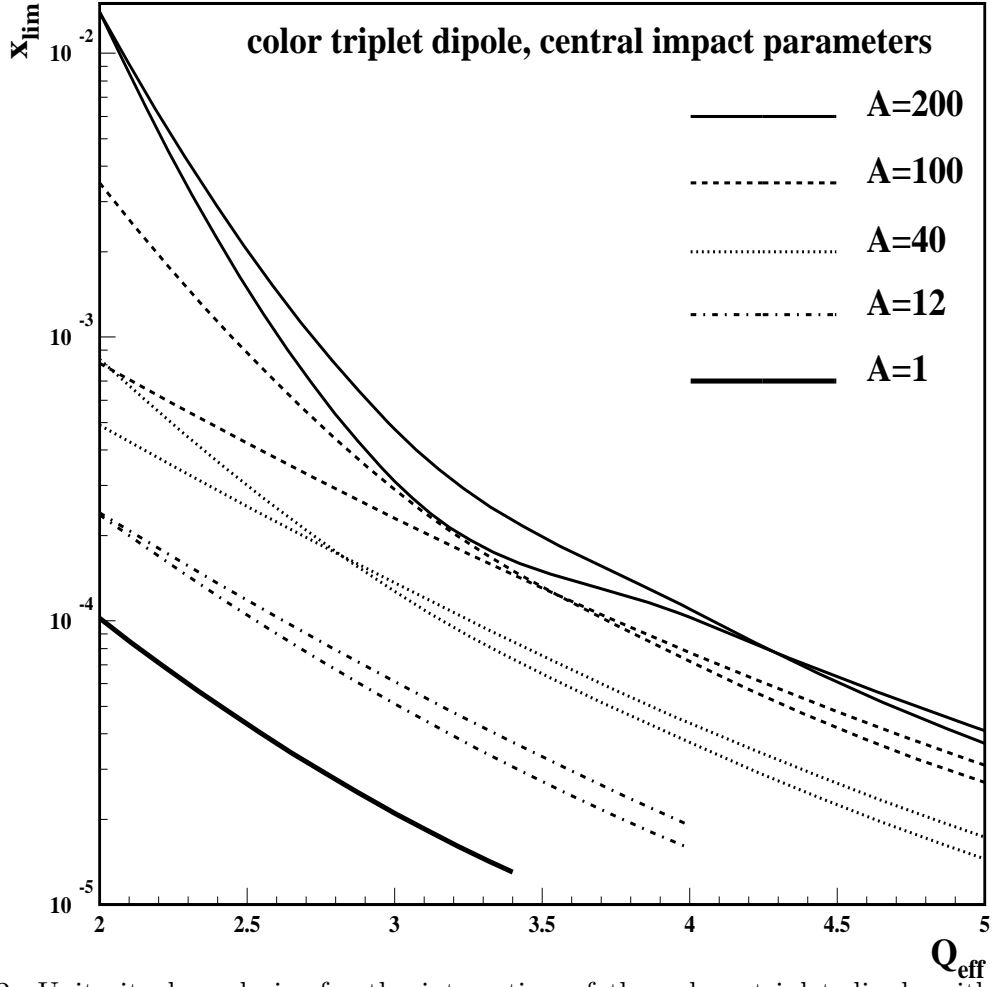}
\caption{Unitarity boundaries for the interaction of the colour 
triplet dipole with nuclei at central impact parameters. Regions to the left of each curve are 
prohibited by the unitarity bound of (\ref{eq:nuc:unit:uc:3}).
Two sets of curves are given for each nucleus correspond to two different 
models of leading twist shadowing (as discussed in subsection \ref{sec:nuc:shad}).}
\label{fig:nuc:unit:1}
\end{center}
\end{figure}

\begin{figure}
\begin{center}
\epsfig{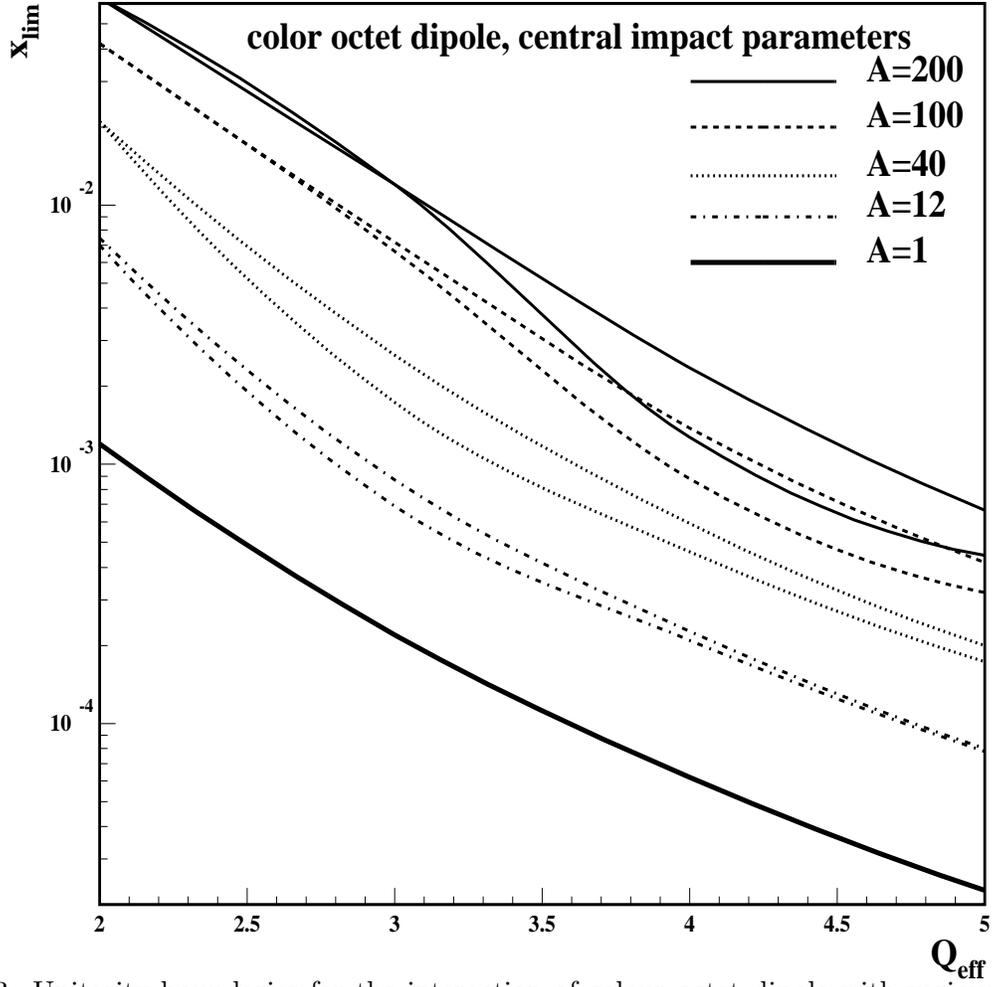}
\caption{Unitarity boundaries for the interaction of
colour octet dipole with various nuclei at central impact parameters.
Regions to the left of each curve are prohibited by the unitarity bound of (\ref{eq:nuc:unit:uc:4}).
Two sets of curves are given for each nucleus correspond to two different 
models of leading twist shadowing (as discussed in subsection \ref{sec:nuc:shad}).}

\label{fig:nuc:unit:2}
\end{center}
\end{figure}

The unitarity constraints are even more stringent for DIS on nuclei at central
impact parameters $b$, $b \leq R_A$.
This is essentially equivalent to scattering off a cylinder of
the length $2 R_A$ oriented along the reaction axis.
Obviously, in this case, edge effects are suppressed and
one also gains an additional factor of $\sim 1.5$ on
the left hand side of (\ref{eq:nuc:unit:uc:2}) due to
the increased density of nucleons.
The nuclear gluon distribution at a given impact parameter $b$
is introduced as \cite{FLS90bnl}
\begin{equation}
xG^{A}(x,Q^2,b) \equiv A~xG^N(x,Q^2)~f^{A}(x,Q^2,b)~T^{A}(b) \ ,
\label{eq:nuc:unit:uc:30}
\end{equation}
where the function $f^{A} (x,Q^2,b)$ describes the amount
of nuclear shadowing,
$T^{A}(b)=\int^{\infty}_{-\infty} dz \rho^{A} (b,z)$
and $\int d^2 b T^{A} (b)=1$. Note that nuclear shadowing is
larger at central impact parameters then in the situation
when one averages over all impact parameters.
Now, the unitarity constraints for DIS on nuclei at central impact parameters
immediately follow from (\ref{eq:nuc:unit:uc:2}) for colour triplet
\begin{equation}
1.5 \times \frac{\pi^2}{3} r^2 \left[ x
G^{A}(x, {\bar Q^2},b=0) \right] \alpha_{s}({\bar Q^2})
\lsim \pi R_A^2\ \, ,
\label{eq:nuc:unit:uc:3}
\end{equation}
and colour octet dipoles
\begin{equation}
1.5 \times\frac{3\pi^2}{4} r^2 \left[ x
G^{A}(x,{\bar Q^2},b=0) \right] \alpha_{s}({\bar Q^2})
\lsim \pi R_A^2 \ \, .
\label{eq:nuc:unit:uc:4}
\end{equation}
The kinematical regions prohibited by these unitarity constraints (defined by
$x < x_{\mbox{{\footnotesize lim}}}, {\bar Q} < Q_{\mbox{{\footnotesize eff}}}$) lie to the left of the curves
in Figs.~\ref{fig:nuc:unit:1} and \ref{fig:nuc:unit:2} (on the boundary
${\bar Q} \equiv Q_{\mbox{{\footnotesize eff}}}, x
\equiv x_{\mbox{{\footnotesize lim}}}$). For the nucleon gluon density
we used the CTEQ4L \cite{cteq4l} parameterization evaluated
at the scale $Q^2 =4$~GeV$^2$.
For each nucleus, we present scenarios with the highest and
lowest shadowing (see Sect.~\ref{sec:nuc:shad}). The curves with more shadowing lie below 
the ones with less shadowing, for all $Q^2$.
To illustrate the trends given by (\ref{eq:nuc:unit:uc:3}), the curves
are extended to the region $x\ge 1 /(4 R_A m_N)$ where,
strictly speaking, the unitarity
constraints (of (\ref{eq:nuc:unit:uc:3}) and (\ref{eq:nuc:unit:uc:4}))
should not be directly applied.

As one can see from Figs.~\ref{fig:nuc:unit:1} and \ref{fig:nuc:unit:2},
effects associated with the unitarity constraints are expected, in a wide range
of $x$ and $Q^2$, to be covered by THERA. Regardless of
the nature of such effects,
strong modifications of the gluon field in heavy nuclei
(as compared to the incoherent sum of the nucleon fields)
appear to be unavoidable.

In summary, in the search for the BBL we observe
that employing nuclear targets
allow us to gain substantially in the region where
higher twist effects become important.
However, the presence of the leading twist nuclear shadowing
reduces the magnitude of this gain.
It is very important that in a wide range $Q^2$ the unitarity limit is reached 
at relatively large $x$ so that $\ln Q^2/\Lambda_{\mbox{{\scriptsize QCD}}}^2 $
is comparable to  $\ln x_0/x $ (where $x_0\sim 0.05$ is the starting point 
for the gluon emission in the $\ln 1/x$ evolution).
Hence, the diffusion to the small transverse momenta is likely to be a correction. 
Therefore non-perturbative QCD, with a large coupling constant, is unlikely to be relevant for the taming of the 
structure functions.

\section{Nuclear shadowing and diffraction}
\label{sec:nuc:shad}

A long time ago Gribov \cite{Gribov691} established an
unambiguous connection between
the cross section of small-$t$ diffraction of a hadron off
a nucleon and the amount
of shadowing in the interaction of the same hadron with a
nucleus, for the limit in which
only two nucleons of the nucleus are involved. Applying
Gribov's formulae to describe
photon-deuteron scattering, the effect of nuclear
shadowing for the total cross
section can be expressed
\footnote{
The contribution of the real part of the
diffractive scattering amplitude was neglected
in \cite{Gribov691} since it
 was assumed that the total cross section is
energy-independent.} as \cite{FSAGK}
\begin{equation}
\sigma_{\mbox{{\footnotesize shad}}}={\sigma_{\mbox{{\footnotesize tot}}} (eD)- 2\sigma_{\mbox{{\footnotesize tot}}} (ep)
\over \sigma(ep)}={(1-\lambda^2)\over (1+\lambda^2)}
{ {d\sigma_{\mbox{{\footnotesize diff}}} (ep) \over dt}
\big|_{t=0}
\over
\sigma_{\mbox{{\footnotesize tot}}} (ep)}{1 \over 8 \pi R_D^2} \ ,
\label{eq:nuc:shad:1}
\end{equation}
where $\lambda$ is the ratio of real to imaginary parts of
the amplitude for diffractive
DIS, and $R_D$ is the deuteron radius. For small $x$,
$\lambda$ may be as large as $0.2 \div 0.3$,
which results in $(1-\lambda^2)/(1+\lambda^2) \sim 0.8 \div 0.9$.
For simplicity, we have neglected the longitudinal nuclear form factor
(which leads to a cutoff of the integral over the produced masses, 
cf. e.g., (\ref{eq:nuc:black_nuc:bbl:8}))

It is worth emphasising that (\ref{eq:nuc:shad:1}) does
not require the dominance of the leading twist in diffraction.
The only assumption, which is
due to a small binding energy of the deuteron
and which is also known to work very well in calculations of
hadron-nucleus total and elastic cross sections, is that the
nucleus can be described as a multi-nucleon system
\footnote{The condition that the
matrix element $<\!A|T[J_{\mu}(y),J_{\nu}(0)]|A>$
(which defines the nuclear structure function and total cross section)
involves only nucleonic
initial and final states is not so obvious in the infinite momentum frame.
However, it is implemented in most of the light-cone models
\cite{Mueller-Qiu,McLerran}.}.
Under these natural assumptions, one is essentially not sensitive to
any details of the nuclear structure, such as short-range nucleon
correlations, etc.

\subsection{Inclusive diffraction at HERA and predictions
for nuclear parton densities}
\label{sec:nuc:shad:inclusive}

The Gribov approximation has been applied to the
description of shadowing in nuclear DIS
for a long time. The first calculations were performed in
\cite{FS88,FS89} before the advent of HERA
and, hence, required modelling of diffraction in DIS. This modelling
was based on the QCD extension
of the Bjorken aligned jet model \cite{FS88}, and produced
a reasonable description of the NMC
data \cite{NMC}. More recently, an explicit use of
the HERA diffractive data allowed an essentially
parameter-free description of these data \cite{Capella} to be provided.

Two important features of the HERA inclusive diffractive
data \cite{H1diff,zdiff}
are the approximate Bjorken scaling for the diffractive parton densities and a
weak dependence of the total probability of diffraction
$P_{\mbox{{\footnotesize diff}}}$ on $Q^2$:
$P_{\mbox{{\footnotesize diff}}} \sim 10\%$ at $Q^2 \geq 4$~GeV$^2$.
The first observation is in line with the Collins factorization
theorem \cite{Collins}
which states that in the Bjorken limit,
the diffractive
structure functions $f_j^D(\beta,Q^2,x_{\Pomeron},t)$ satisfy
the
DGLAP evolution equations ($\beta=x/x_{\Pomeron} \approx Q^2/(Q^2+M^2)$ and
$x_{\Pomeron} \approx (M^2 + Q^2) / 2q\cdot(p-p')$ is the
fraction of the proton's momentum
 carried by the diffractive exchange).

A relatively small value of the probability of diffraction
(as compared to the case of $\pi N$ scattering)
indicates that the average strength of the
interaction leading to diffraction is correspondingly small.
To characterise this strength
it is instructive to treat diffraction in the
$S$-channel picture
using the eigenstates
of the scattering matrix \cite{eigen}.
Such an approach is complementary to the picture of the factorization theorem.
The use of the optical theorem leads to the following
definition of the effective strength of the interaction 
(for diffraction in nucleon collisions and shadowing in nuclear collisions):
\begin{equation}
\sigma_{\mbox{{\footnotesize eff}}}^j(x,Q^2)\equiv 16 \pi {{ d \sigma^j_{\mbox{{\footnotesize diff}}} \over dt} \big|_{t=0}
\over \sigma^j_{\mbox{{\footnotesize tot}}} (x,Q^2)} \ .
\label{eq:nuc:shad:inclusive:1}
\end{equation}
Here the superscript $j$ indicates which hard parton is involved in the
elementary hard process of the cross section $\sigma_{\mbox{{\footnotesize tot}}}(x,Q^2)$.
This equation allows the average cross section for configurations which contribute to
quark-induced and gluon-induced diffraction to be extracted.

Recently a number of phenomenological analyses of HERA
inclusive diffractive and diffractive jet
production data were performed within the leading twist
approximation for hard diffraction.
We have compared several of these analyses
\cite{H1diff,ACWT,HS} and found that they lead to
fairly similar values of $\sigma_{\mbox{{\footnotesize eff}}}^j$, especially if a matching at large
diffractive masses to the soft factorization was
implemented (for details see \cite{FGMS01}).

\begin{figure}
\begin{center}
\epsfig{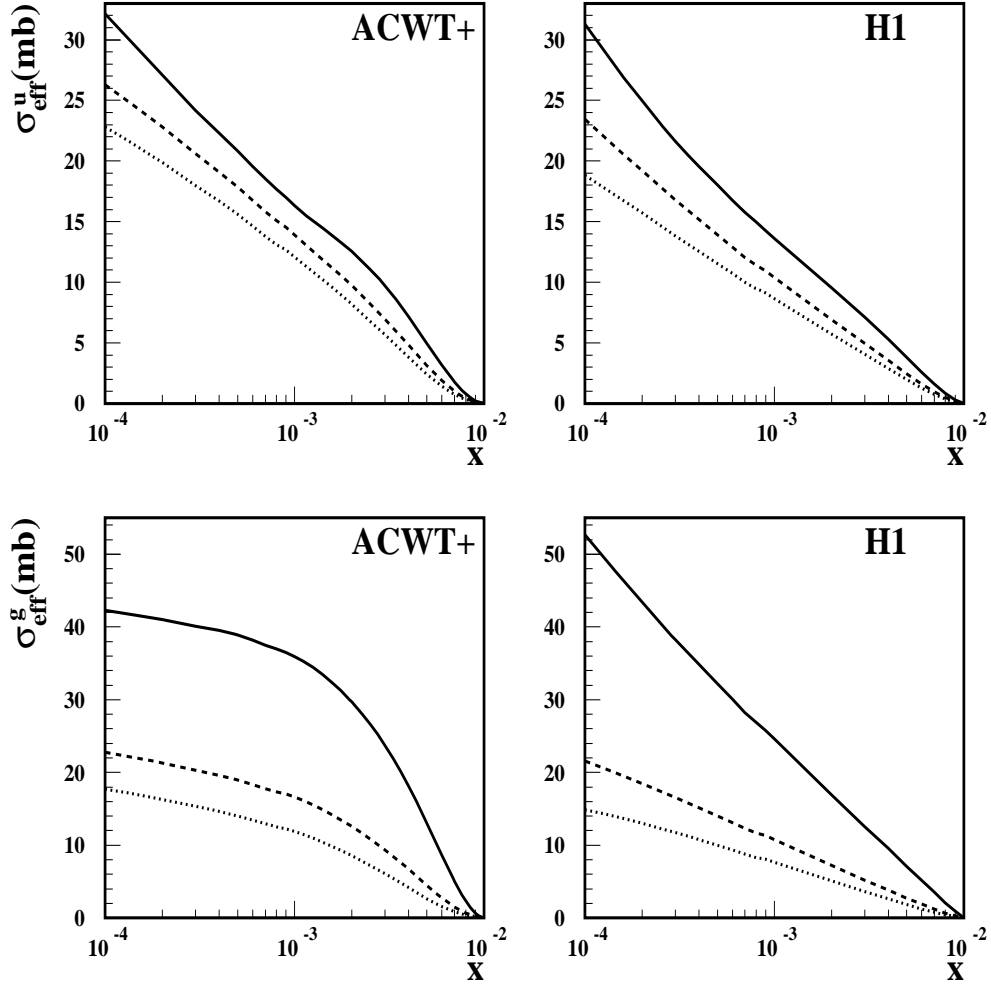}
\caption{The effective cross section for nuclear shadowing (or nucleon diffraction) 
of (\ref{eq:nuc:shad:inclusive:1}), $\sigma_{\mbox{{\footnotesize eff}}}^j$, 
for a parton of type $j = u,g$, as a function of Bjorken $x$ at several values of $Q^2$. 
The solid curve is for $Q =2$~GeV, the dashed curve is for $Q =5$~GeV, the dotted one is for $Q=10$~GeV. 
Two representative \protect\cite{HS} sets of diffractive parton distributions have been used in the curves 
labelled by ``ACWT+'' \protect\cite{ACWT} and ``H1'' \protect\cite{H1diff} (see text).}
\label{fig:nuc:shad:inclusive:1}
\end{center}
\end{figure}

The results for $\sigma_{\mbox{{\footnotesize eff}}}^j$ for up quarks and
gluons are presented in Fig.~\ref{fig:nuc:shad:inclusive:1}. We
used results of two analyses of the HERA diffractive data (by the H1 collaboration \cite{H1diff}, labelled by ``H1'' in the figure, and  by Alvero {\it et al.} \cite{ACWT}, labelled by ``ACWT+''). The
diffractive parton densities of \cite{ACWT} were extended to the region of small $\beta$ by requiring
consistency with the soft factorization theorem. Three curves (solid, dashed and dotted) represent
the $Q^2$-evolution of $\sigma_{\mbox{{\footnotesize eff}}}^j$. While the $Q^2$-evolution
is not very significant for $\sigma_{\mbox{{\footnotesize eff}}}^u$ of the up quark,
it decreases $\sigma^{g}_{\mbox{{\footnotesize eff}}}$ rather rapidly for gluons. This rapid change
is explained by large scaling violations for $xG^N (x,Q^2)$.

\begin{figure}
\begin{center}
\epsfig{file=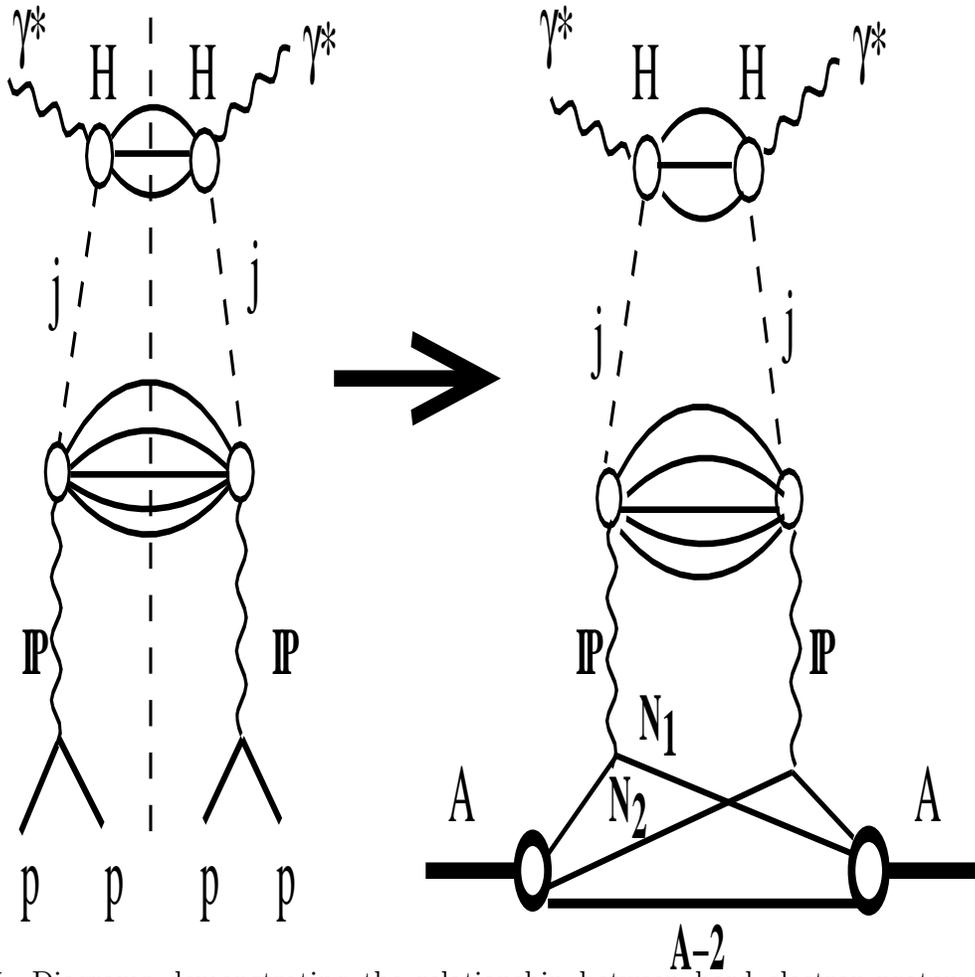,width=13cm,height=13cm}
\caption{Diagrams demonstrating the relationship between hard electron-proton diffraction, 
involving a parton of type $j$, and nuclear shadowing in inclusive DIS. 
Here the Pomeron symbol, $\Pomeron$, merely represents a generic label for vacuum exchange.}
\label{fig:nuc:shad:inclusive:2}
\end{center}
\end{figure}

Combining the Gribov theory with
the Collins factorization theorem
and comparing the QCD diagrams for hard
diffraction and for nuclear shadowing
arising from the
 scattering off two nucleons (see Fig.~\ref{fig:nuc:shad:inclusive:2}),
one can prove \cite{FS992}
 that, in the low thickness limit, the leading twist
nuclear shadowing is unambiguously expressed through the diffractive
parton densities
$f_j^D({x / x_{\Pomeron}},Q^2,x_{\Pomeron},t)$
of $ep$ scattering:
\begin{eqnarray}
&&f_{j/A}(x,Q^2)/A = f_{j/N}(x,Q^2)
 -{1 \over 2}\frac{1-\lambda^2}{1+\lambda^2}
\int d^2b\int_{-\infty}^{\infty}dz_1\int_{z_1}^{\infty} dz_2
\int_x^{x_0} dx_{\Pomeron} \nonumber \\
&&\times
\rho_A(b,z_1)\rho_A(b,z_2)
\cos(x_{\Pomeron}m_N(z_1-z_2))
f^{D}_{j/N} \left(\beta, Q^2,x_{\Pomeron},t\right) \big|_{k_t^2=0} \, ,
\label{eq:nuc:shad:inclusive:2}
\end{eqnarray}
where $f_{j/A}(x,Q^2)$ and $f_{j/N}(x,Q^2)$ are the
inclusive parton densities,
$\rho_A$ is the nucleon density in the nucleus. At moderately 
small values of $x$, one should also add a term related to the
longitudinal distances comparable to the inter-nucleon
distances in the nucleus. This additional term can be
evaluated using information on
the enhancement of the gluon and valence quark parton densities at $x \sim 0.1$
at the initial scale $Q_{0}^2$.
This would slightly diminish
 nuclear shadowing at higher $Q^2$ via the $Q^2$ evolution.
To summarise the results of (\ref{eq:nuc:shad:inclusive:2}),
the nuclear shadowing effect given by $(1-f_{j/A}(x,Q^2)/Af_{j/N}(x,Q^2))$,
is proportional to $\sigma^j_{\mbox{{\footnotesize eff}}} (x,Q^2)$.

\begin{figure}
\begin{center}
\epsfig{file=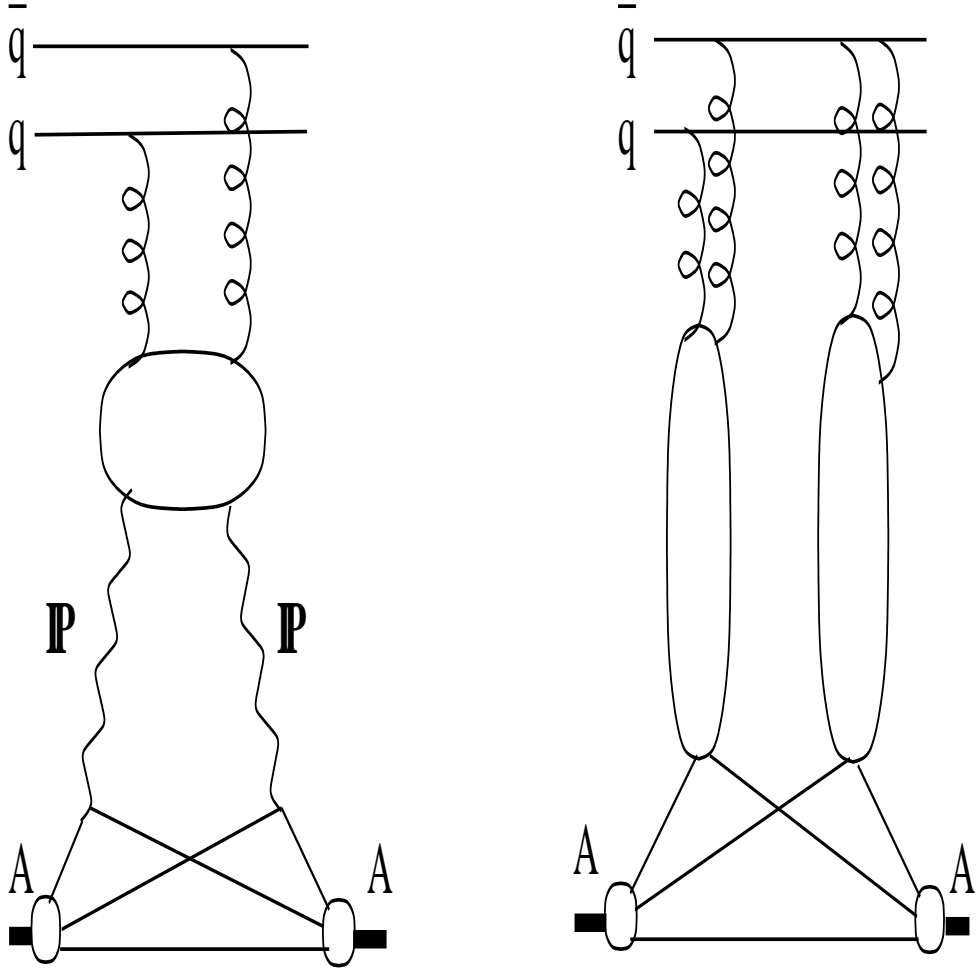,width=13cm,height=13cm}
\caption{Typical diagrams for the leading twist (left) and
eikonal (right) models for nuclear shadowing involving a small dipole and two nucleons. The curly lines represent gluons.
In the diagram on the right the blob and its attached legs represent the gluon distribution of the nucleon.
The Pomeron symbol, $\Pomeron$, in the diagram on the left merely represents a generic label for vacuum exchange.}
\label{fig:nuc:shad:inclusive:3}
\end{center}
\end{figure}

For $N \geq 3$ nucleons, we need to establish which
configurations in the photon wavefunction
dominate in diffraction. There are two extreme alternatives, i.e.
a dominance of either
hadronic-size configurations (as indicated by the aligned jet model)
or of small size
($\sim 1/Q$) fluctuations. Figure~\ref{fig:nuc:shad:inclusive:3}
represents the corresponding diagrams
(contributing to nuclear shadowing).

One can determine which of these extremes is closer to reality by comparing
the effective cross section characterising diffraction
(\ref{eq:nuc:shad:inclusive:1})
with the cross section,
$\sigma_{\mbox{{\scriptsize pQCD}}} (\dperp^2,x)$, for the double
interaction of a dipole with a small and fixed diameter
$\dperp \sim 1/Q$, within the
eikonal approach, given by (\ref{eq:nuc:intr:eqshat}).
Thus, for the $x$-range studied at HERA at
$Q^2\geq 4$~GeV$^2$ ($x \sim 0.001$) ,
large-size configurations dominate, while the small dipoles
contribute little to the bulk of the
inclusive diffractive cross section. This can also be seen
from a comparison of the amount of
nuclear shadowing in $F_L^A$ calculated within the leading
twist and eikonal approaches.
This is consistent with the experimental observation that
the $t$-slope of the inclusive diffraction
($B \sim 7$~GeV$^{-2}$) is significantly larger than for
the processes where small-size dipole
dominate ($B \sim 4.5$~GeV$^{-2}$). At lower $x \sim 10^{-4}$, small
dipoles may become much more
important due to the increase of $\hat{\sigma}_{\mbox{{\scriptsize pQCD}}}$
due to the QCD evolution.

\begin{figure}
\begin{center}
\epsfig{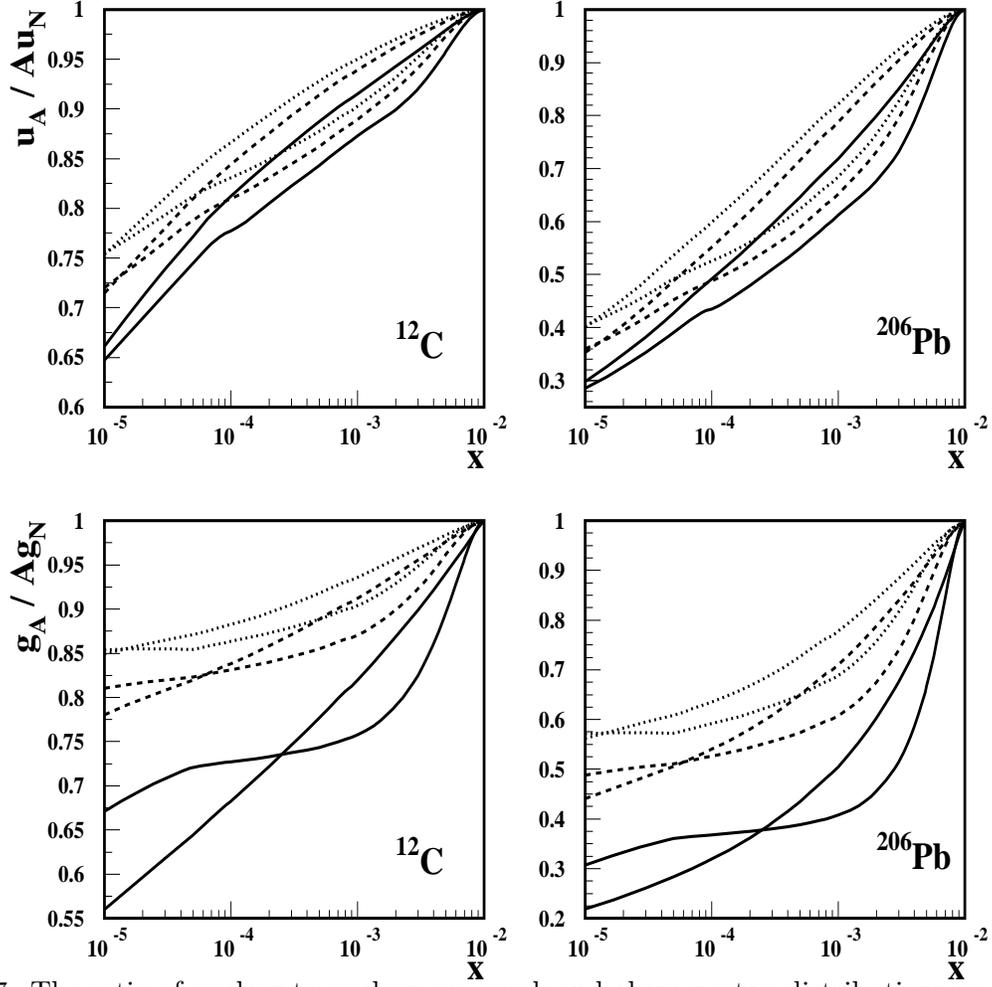}
\caption{The ratio of nuclear to nucleon up-quark and gluon parton distributions as a function of $x$, scaled by 
nucleon number, $A$. Two representative \protect\cite{HS} sets of diffractive parton densities, ``ACWT+'' \protect\cite{ACWT}
and ``H1'' \protect\cite{H1diff}, are used to calculate nuclear shadowing 
(see text) for each nucleon and taken together they give 
an indication of the spread of theoretical predictions.
The solid, dashed, and dotted curves are for $Q =2, 5, 10$~GeV, respectively.}
\label{fig:nuc:shad:inclusive:4}
\end{center}
\end{figure}

It is well known that total, elastic and inelastic diffractive
cross sections for the
interactions of hadrons with nuclei are quantitatively
well-described within the scattering
eigenstate approximation \cite{eigen}, which is a generalization
of the eikonal approximation.
Since, at $Q^2 =4$~GeV$^2$, relatively soft interactions give the dominant
contribution to diffraction, interactions with $N \geq 3$
nucleons can be considered using a generalised eikonal model with
an effective cross section given by (\ref{eq:nuc:shad:inclusive:1}).
We found that, due to a relatively small value of $\sigma_{\mbox{{\footnotesize eff}}}^q$, the
amount of nuclear shadowing for $F^{A}_{2}$ is relatively
modest (as compared to the nuclear gluon distribution) and is practically independent
of fluctuations of the effective cross section (when $\sigma_{\mbox{{\footnotesize eff}}}$ is
kept fixed) \cite{FS991}.
The $Q^2$-dependence of shadowing is taken into account via
the DGLAP evolution equations.
Recently, we performed an extensive comparison of different parametrisations
of the diffractive structure functions \cite{FGMS01}. The results
will update the analysis
\cite{FS992} to include a range of modern parameterizations of
the quark and gluon diffractive parton
densities. Using the effective cross sections presented
in Fig.~\ref{fig:nuc:shad:inclusive:1} and
generalising (\ref{eq:nuc:shad:inclusive:2}) to include
the rescattering terms with $N \geq 3$ nucleons,
we are able to produce predictions for up quark and gluon
parton distributions for a number of nuclei.
As an example Fig.~\ref{fig:nuc:shad:inclusive:4} represents
the ratio of nuclear to nucleon up quark
and gluon parton distributions for nuclei of
$^{12}C$ (carbon) and $^{206}Pb$ (lead). Note that since
the models we used provide good fits to the HERA diffractive data
for $F_2^D (x, Q^2, x_{\Pomeron})$,
they effectively take into account higher twist terms, if
they are present in the data.

Hence, we conclude that combining the Gribov theory with the
factorization theorem for inclusive
diffraction and experimental information from HERA
provides reliable predictions for nuclear
parton distributions and, hence, for $F^{A}_{2}$ in
the HERA and THERA experimental ranges.

\subsection{Gluon shadowing}

We explained above that since the
interaction of the
colour octet dipole is enhanced by a factor 9/4
relative to a colour triplet \cite{rad,afs,bhm}, one expects,
for processes sensitive to such configurations, an earlier
onset of the regime where unitarity
effects may become important.

The leading twist mechanism of nuclear shadowing is connected to the
amount of the leading twist diffraction in gluon-induced reactions. 
The larger strength of the perturbative interaction as well as stronger 
non-perturbative interactions up to a scale $\sim 2$~GeV in the gluon channel
suggest that the gluon-induced diffraction should occur with a large probability.

This is consistent with another interesting feature of the current HERA diffractive
data (shared by the diffractive data from the proton colliders), i.e. a 
very important role of the gluons in hard diffraction. In the language of 
the diffractive community, the ``perturbative Pomeron'' is predominantly
built of gluons. To quantify this statement it is convenient to define the probability
of diffraction for a hard probe which couples to a parton $j$ \cite{FS992}:
\begin{equation}
P_j(x,Q^2)=
{\int dt dx_{\Pomeron} ~f_j^D({x/x_{\Pomeron}}, Q^2, x_{\Pomeron},t) \over
f_j(x,Q^2)} \, .
\label{eq:nuc:shad:inclusive:4}
\end{equation}
A large probability of diffraction in the gluon channel:
\begin{eqnarray}
P_g (x \leq 10^{-3}, Q^2=~4~\mbox{GeV}^2)~\sim~0.4 \, , & \nonumber \\
P_g (x \leq 10^{-3}, Q^2=10~\mbox{GeV}^2)~\sim~0.2 \, , & \nonumber \\
\end{eqnarray}
leads via (\ref{eq:nuc:shad:inclusive:4})
to a large $\sigma_{\mbox{{\footnotesize eff}}}^g$ (see Fig.~\ref{fig:nuc:shad:inclusive:1})
and, hence, to
the prediction of {\it a large leading twist shadowing}
for $G^A (x \leq 10^{-2}, Q^2)$
(see Fig.~\ref{fig:nuc:shad:inclusive:4}).
Interactions of small dipoles are more important in the gluon case than in
the quark case.
Nevertheless, if we use the eikonal approximation
(in the same spirit as for $q\bar q$ dipoles)
to estimate the relative importance of the leading and higher twist effects,
we find that up to fairly small $x$ in a wide range of $Q^2$,
the leading twist terms dominate.
Note that in this case one should use
$\sigma_{q {\bar q} N}$ from \cite{mfgs} rescaled
by the factor 9/4 even for the dipole sizes where non-perturbative
effects could be important.
A comparison of Figs.~\ref{fig:nuc:unit:1} and \ref{fig:nuc:unit:2}
suggests that the higher
twist effects appear to be more important in the gluon channel
than in the quark channel.
Also, the gluon induced interactions of a projectile with several nucleons are
much more important and more sensitive to details of the interaction dynamics.
However, with $\sigma_{\mbox{{\footnotesize eff}}}^{g} \sim 40$~mb, a large
leading twist shadowing in the gluon channel
appears to be unavoidable. It seems to be large enough to reduce
strongly the gluon densities.
However, even this strong reduction can not prevent
the violation of unitarity for the interaction of colour octet
systems with heavy nuclei, as
can be seen from Fig.~\ref{fig:nuc:unit:2}. For scattering
at central impact parameters,
this may already occur at $Q \sim 2$~GeV for 
the whole $x$ range, $x \le 1 /(4 m_NR_A)$.

\subsection{Shadowing in the interaction of small dipoles with nuclei}
\label{sec:nuc:shad:small}

For several small-$x$ processes, we can probe the
interactions of a small colour dipole with the nucleus directly.
These include $\sigma_L (x,Q^2)$, $\sigma (\gamma+A \to J/\psi +A)$
and $\sigma(\gamma^{\ast}_L+A \to \rho +A)$.

We argued above that the eikonal approximation
for fixed small $\dperp$ leads to much weaker
absorption effects than the leading twist
shadowing. The eikonal approach is also inconsistent
with the QCD factorization theorem for the production
of vector mesons \cite{cfs} which leads at small $x$ to \cite{brod}
\begin{eqnarray}
{{d\sigma\over dt}(\gamma^{\ast}_{L} A \to V A)\big\vert_{t=0}\over
{d\sigma\over dt}(\gamma^{\ast}_{L} N \to V N)\big\vert_{t=0}}
\approx \left [{F^{A}_{L} (x,Q) \over F^{N}_{L} (x,Q)}\right ]^2
\approx \left [ {xG^A (x,Q) \over xG^N (x,Q)} \right ]^2 \ .
\label{eq:nuc:shad:small:1}
\end{eqnarray}

\begin{figure}
\begin{center}
\epsfig{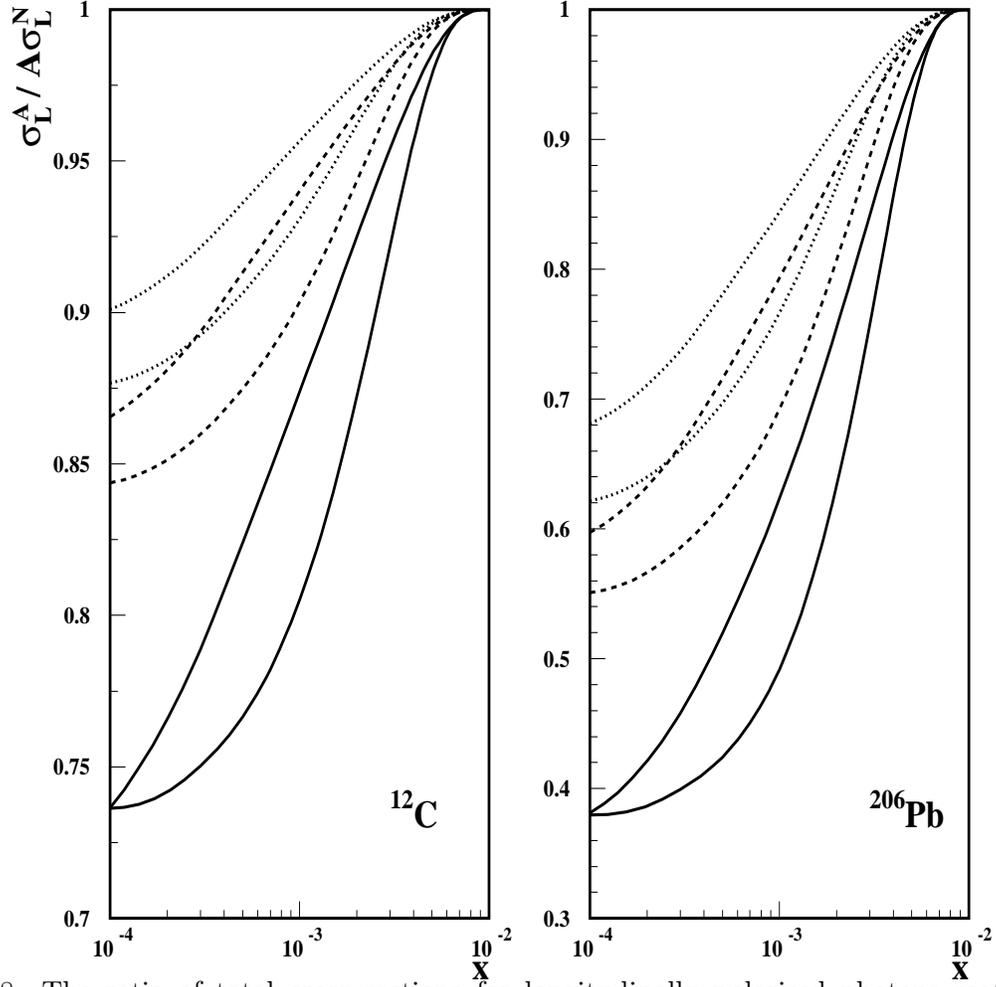}
\caption{The ratio of total cross sections for longitudinally polarised photons scattering on 
 nuclear and nucleon targets, $\sigma^{A}_{L}/(A\sigma^N_{L})$, as function of $x$ for Carbon and Lead.
Two representative \protect\cite{HS} sets of diffractive parton densities are used, 
``ACWT+'' \protect\cite{ACWT}  and ``H1'' \protect\cite{H1diff}, to calculate nuclear shadowing. 
The solid, dashed and dotted curves are for $Q =2,5,10$~GeV, respectively.}
\label{fig:nuc:shad:small:1}
\end{center}
\end{figure}

Another observable sensitive to the difference between the leading twist and eikonal approaches is
$F^{A}_{L}$. It can be expressed via  nuclear parton  distributions (we consider the leading order in $\alpha_{s}$ expression):
\begin{equation}
F^{A}_{L}(x,Q^2)=\frac{2 \alpha_{s}(Q^2)}{\pi} \int^{1}_{x} \frac{dy}{y} (x/y)^2 \Big(\sum^{n_{f}}_{i=1} e_{i}^2   (1-x/y) y G^{A}(y,Q^2)+\frac{2}{3}F_{2}^{A}(y,Q^2)\Big) \ .
\label{eq:nuc:shad:small:2}
\end{equation}
The amount of nuclear shadowing for $\sigma^{A}_{L}$ can be expressed by the ratio
\begin{equation}
\frac{F^{A}_{L}}{A F^N_{L}}=  \frac{\int^{1}_{x} \frac{dy}{y} (x/y)^2 \Big((1-x/y)~\sum_i e_i^2~yG^{A}(y,Q^2)+ 2 F_{2}^{A}(y,Q^2)/3 \Big)}{A \int^{1}_{x} \frac{dy}{y} (x/y)^2 \Big((1-x/y) \sum_{i} e_i^2~yG^N (y,Q^2)+2 F^N_{2}(y,Q^2)/3 \Big)} \ .
\label{eq:nuc:shad:small:3}
\end{equation}
Results using (\ref{eq:nuc:shad:small:3}) are presented in Fig.~\ref{fig:nuc:shad:small:1}
for the same parameterizations as in the previous figures.

\begin{figure}
\begin{center}
\epsfig{file=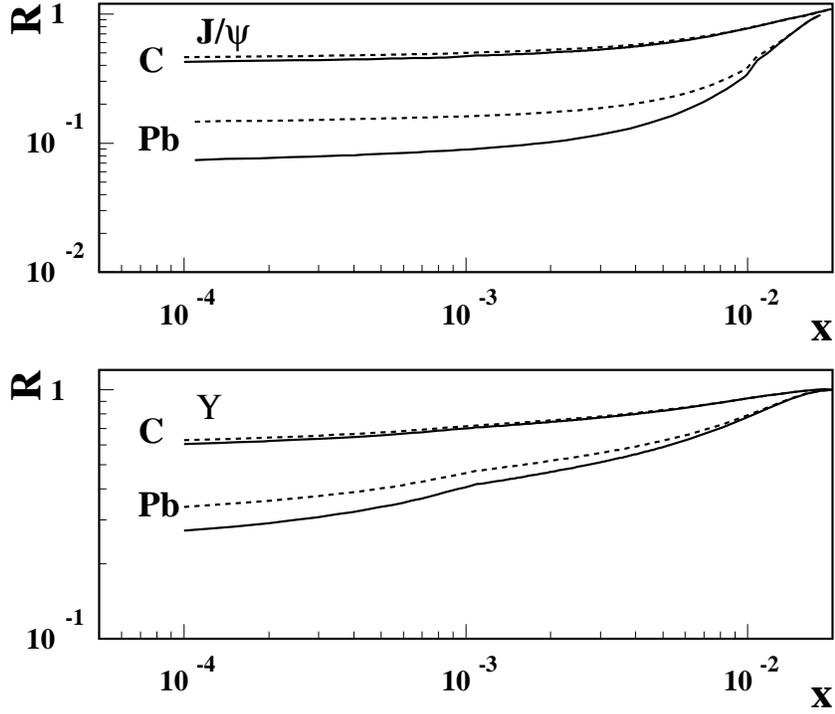,width=13cm,height=13cm}
\vspace{-1cm}
\caption{The 
colour opacity effect for the ratio, $R$, 
of the coherent production of $J/\psi$ and $\Upsilon$ from Carbon and Lead,
normalized to the value of this ratio at $x=0.02$, calculated in the leading twist models of gluon
shadowing \protect\cite{FS992} both  with (dashed curves) and without (solid curves) taking the fluctuations of the cross section into account.}
\label{fig:nuc:shad:psi}
\end{center}
\end{figure}

The impact-parameter eikonal approximation
for $x \ll 1 / (4 m_N R_A$) (where finite coherence length effects, which cannot be unambiguously
treated in the eikonal approximation, can be safely neglected),
leads to substantially smaller
shadowing for the interaction of small dipoles
(see Fig.~(15) of \cite{fks1}).
The  cleanest way to observe this effect would be to study coherent
$J/\psi$ production. THERA would be able to cover a broad range of
$x= (M_{J/\psi}^2 + Q^2)/W^2$ starting from the region of colour transparency
(amplitude proportional to nucleon number $A$) and extending to the
colour opacity regime where there is a strong reduction of the amplitude
(see Fig.~\ref{fig:nuc:shad:psi}).

\subsection{Total cross section of coherent inclusive diffraction}

\begin{figure}
\begin{center}
\epsfig{file=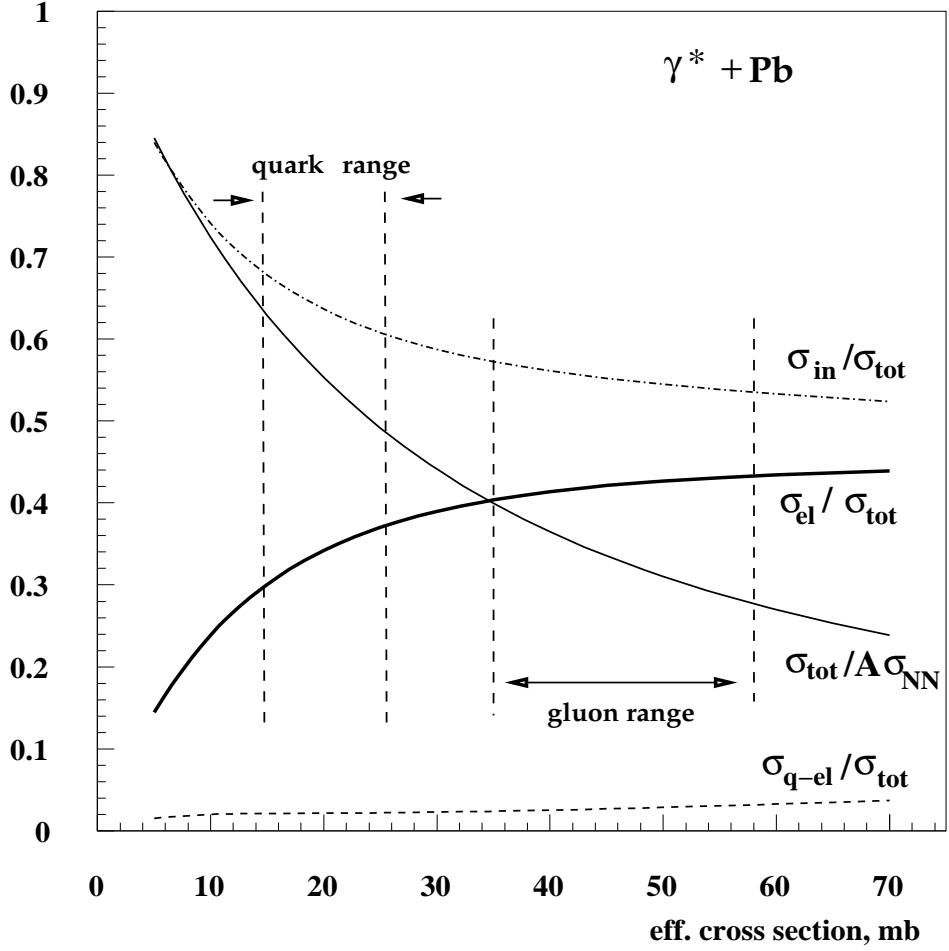,width=13cm,height=13cm}
\caption{Fractions of the total cross section, given by inelastic scattering ($in$), 
coherent ($el$) and incoherent ($q-el$) diffractive scattering cross sections as a function of the 
average interaction strength for quarks and gluons ($\sigma^{j}_{\mbox{{\footnotesize eff}}}$ of (\ref{eq:nuc:shad:inclusive:1}), cf. Fig. 
\ref{fig:nuc:shad:inclusive:1}) for scattering off Lead.
}
\label{fig:nuc:shad:gap}
\end{center}
\end{figure}

There is a deep connection between shadowing and the phenomenon of
diffractive scattering off nuclei. The simplest way to
investigate this connection is to apply
the AGK cutting rules \cite{AGK}.
Several processes contribute to diffraction on nuclei:
(i) coherent diffraction in which the nucleus remains intact,
(ii) break-up of the nucleus (without producing of hadrons in the nucleus fragmentation
region), (iii) rapidity gap events (with hadron production in the nucleus fragmentation region).
In \cite{FSAGK} we found that for $x\leq 3 \cdot 10^{-3}, Q^2\ge 4$~GeV$^2$, the
fraction of the DIS events with rapidity gaps reaches about 30-40\%
for heavy nuclei, with a fraction of the events of type (iii) decreasing rapidly with $A$.
Recently, together with M.~Zhalov one of us (MS) investigated
the dependence of the fraction of the events due to coherent diffraction and due to the break-up of the nucleus
on the strength of the interaction, $\sigma_{\mbox{{\footnotesize eff}}}^j$.
We found that this fraction increases with $\sigma_{\mbox{{\footnotesize eff}}}^j$ rather slowly.
Thus, it is not sensitive to fluctuations of $\sigma_{\mbox{{\footnotesize eff}}}$.
One can see from Fig.~\ref{fig:nuc:shad:gap} that one
expects a significantly smaller fraction for quark-induced processes ($\approx 35 \%$)
of coherent events than for gluon-induced processes ($\approx 45 \%$).
We also found that the ratio of diffraction with and without nuclear break-up is small (10-20\%)
in a wide range of nuclei, and weakly depends on $\sigma_{\mbox{{\footnotesize eff}}}$.
Hence, it would be pretty straightforward to extract coherent diffraction
by simply using anti-coincidence with a forward neutron detector,
especially in the case of heavy nuclei (see discussion in \cite{STZ}).

It is worth emphasising that the proximity of $\sigma_{\mbox{{\footnotesize el}}}/\sigma_{\mbox{{\footnotesize tot}}}$ to 1/2 does
not necessarily imply the proximity of the BBL in the sense of hard physics (it may also occur
in the soft aligned-jet type scenario). A key distinction in this regard is in the dominance of the dijet
production, broadening of the $p_t$ distributions, etc, as discussed in Sect.~\ref{sec:nuc:black_nuc}.

Other manifestations of leading twist shadowing include a strong $A$-dependence
of the fluctuations of the central multiplicity distribution of the produced hadrons \cite{FSAGK},
and strong modifications of the leading hadron spectrum \cite{FS992} for rapidities
where the diffractive contribution is small. This is also in marked contrast to the black body
limit, where major modifications occur already at the highest rapidities.

\section{Experimental considerations}

\label{sub:exp}

The expected nuclear effects in the small-$x$ region are large enough so that
measurements of the ratios of the corresponding quantities
for electron-nucleus
and electron-nucleon scattering with an accuracy of only
a few percent will be sufficient.
Hence, the luminosities required for the first run of
measurements are rather modest.
The estimates of \cite{report} indicate that luminosities
of about $1$~pb$^{-1}$
per nucleus would be sufficient for measurements of a
number of important inclusive
observables which include ratios of the
structure functions, the $A$-dependence of
global features of diffractive cross sections, ratios of the leading particle
spectra and fluctuations of the multiplicity in the central rapidity
range. Luminosities
of the order of $10$~pb$^{-1}$ per nucleus will be necessary to measure
gluon densities via 2 + 1 jet events directly, to observe the
pattern of colour opacity in exclusive vector
meson production and to establish
differential diffractive parton densities of nuclei.

Requirements for the detector are practically the same as for the
study of small-$x$ electron-proton scattering. The only additional requirement
is for forward neutron detectors with specifications similar to that for
the ZEUS forward neutron calorimeter (FNC). These will be needed to
trigger on central inelastic collisions by selecting events where a larger than average
number of neutrons is produced. Note that for small $x$ collisions at central impact
parameters the virtual photon interacts inelastically with $\sim A^{1/3}$ nucleons.
On the basis of an analysis \cite{STZ} of E-665 data on the production of soft neutrons at
$x \sim 0.03$ \cite{E665}, we estimate that the average number of soft neutrons
produced in the central collisions will be $ \sim 20$ (as compared to $\sim
4$ for peripheral inelastic collisions). Overall, a strong correlation between the number
produced nucleons and centrality of the event is expected at small $x$ (as is also
the case for hadron-nucleus scattering). As we explained above,
such detectors would allow the breakdown of leading twist QCD to be studied
in runs with a single heavy nucleus.

A FNC-type neutron calorimeter would have nearly
$100\%$ acceptance for the neutrons from nuclear break-up. Since the average
number of such neutrons exceeds one for $A\ge16$, a FNC-type detector would
allow an effective tagging of nuclear break-up. Hence, it would also
simplify studies of diffractive processes with nuclei. The separation of the
three classes of diffractive events (coherent scattering, nuclear break-up producing
nuclear fragments (mostly neutrons and protons) in the forward direction, and diffractive
dissociation with meson production in the nucleus fragmentation region from
non-diffractive reactions) is very similar to the selection of diffractive events in the
$ep$ case (see the previous subsection). The contribution of diffractive dissociation to the overall diffractive cross section is significantly smaller than in the $ep$ case.
The experimental signatures for this class of reactions are similar to $ep$ scattering
since the energy flow in the forward direction is expected to have almost the same topology.
Calorimetric coverage down to $\theta \leq 1^{0}$ in the forward region will allow for the detection of most dissociative reactions.
Nuclear break-up is expected to constitute about $10\%$ of coherent diffraction for
a wide range of model parameters (see Fig.~\ref{fig:nuc:shad:gap}). In the case of
coherent exclusive processes, such as coherent vector meson production, detection of
the process will be further simplified since the average
transverse momenta of coherently
produced vector mesons are $p_t \le 2 / R_A^2$, which is much smaller than for
exclusive processes with protons.

\section{Conclusions}

\label{sec:nuc:conc}

The nuclear program at THERA has a very strong potential for
the discovery of a number of interesting new high energy (small $x$)
phenomena, including quark and gluon shadowing (for kinematics in
which longitudinal distances are much larger than the heavy
nuclear radius), hard coherent diffraction off nuclei and
colour opacity in the production and interaction of small colour singlets. 
Crucially, for the first time, it will provide several effective tools for unambiguously probing 
the black body limit of photon-nuclear collisions, thereby investigating QCD in a new regime
of strong interactions with a small coupling constant. 

We thank A.~H.~Mueller for a useful discussion and 
GIF, ARC, PPARC and DOE for support. LF and MS thank INT for hospitality
during the time this work was completed.

\end{document}